\documentclass[journal=apchd5,manuscript=article]{achemso}

\usepackage[utf8]{inputenc}

\usepackage{chemformula} 
\usepackage[T1]{fontenc} 
\usepackage[hidelinks]{hyperref}
\usepackage{cleveref}
\usepackage{rotating}
\usepackage{xy} 
\usepackage{booktabs}
\usepackage{pifont}
\usepackage{color}
\usepackage{listings}
\usepackage{pdfpages}
\usepackage{array}
\usepackage{algorithm}
\usepackage{algorithmic}
\usepackage{float}
\usepackage{caption}
\usepackage{lastpage}
\usepackage{afterpage}
\usepackage{lipsum}
\usepackage{adjustbox}
\usepackage{chemfig}
\usepackage[version=4]{mhchem}
\usepackage{chemformula}
\usepackage{siunitx}
\usepackage[makeroom]{cancel}
\usepackage{amsmath, amssymb}
\usepackage{multicol}
\usepackage{lipsum}
\usepackage{feynmp}

\author{Oğuzhan Yücel}
\affiliation[METU]
{Department of Physics, Middle East Technical University, Ankara, Turkey}

\author{Serkan Ateş}
\affiliation[IzTech]
{Department of Physics, Izmir Institute of Technology, Izmir, Turkey}

\author{Alpan Bek}
\affiliation[METU]
{Department of Physics, Middle East Technical University, Ankara, Turkey}
\email{bek@metu.edu.tr}

\title[An \textsf{achemso} demo]
  {Single-Photon Nanoantenna with in Situ Fabrication of Plasmonic Ag Nanoparticle at an hBN Defect Center}

\abbreviations{hBN, AgNp, MNP, QE, HBT}
\keywords{Single photon source, hexagonal Boron Nitride (hBN), plasmonics, nanoantenna, dewetting thin films.}

\begin{document}


\begin{abstract}

We present a practical new method for fabricating a coupled single quantum emitter-plasmonic nanoantenna system.
Emission characteristics of a single defect center embedded in hexagonal Boron Nitride (hBN) multilayers is modified using plasmonic nanoantennas. By dewetting thin silver films on hBN multilayers, plasmonic nanoantennas are obtained in a size controlled way with no adverse effects on the defects. A very same single defect center is investigated with and without nanoantenna in order to demonstrate the modification of its emission characteristics.
Based on the initial silver film thickness in dewetting process, on-demand enhancement and quenching effects are observed. For attaining deterministic coupling strengths, an electromagnetic simulation model is employed in the light of experiments. Fluorescence lifetime, radiative and nonradiative emission rate calculations are used for estimating the spatial configuration of the defect-nanoantenna system as well as for confirming the experimental findings. Our approach provides a low-cost and uncomplicated coupling scheme as an alternative to the scanning probe tip antenna technique.
\end{abstract}


Single-photon sources (SPSs) lie at the core of quantum technologies with their purely non-classical light emission.\cite{Lounis2005,santori2002indistinguishable,kiraz2004quantum}
The idea of coupling of an SPS with a cavity originates from modification of spontaneous emission by changing the environment of quantum emitter (QE) which can be used to adjust the performance of an SPS.\cite{Purcell} After first experimental demonstration of modified\cite{Serge1982superradiance} emission from  a single quantum system with a millimeter size cavity and with the advances in microfabrication techniques, microcavities\cite{Vahala2003} came forth due to their comparatively diminutive volumes in the micron range lending potential to integrate into large scale  photonic circuitry.
Furthermore, a plasmonic nanoantenna enables light-matter interaction at the nanoscale. By means of electromagnetic field confinement, ultra-small mode volume (V) engenders significantly high Purcell factor\cite{Chang2006} despite low quality factor ($Q\leq 10^2$).
As achieving a high Purcell factor is possible through attaining a high Q/V ratio, plasmonic nanoantennas\cite{koenderink2010use} are highly attractive choices for this purpose in solid state quantum optics.\par
Purcell factor is an overall indicator of spectral, spatial and polarization match of a quantum emitter with a nanoantenna\cite{giannini2011plasmonic}.
An ideal spectral match requires nanoantenna to have same resonance frequency with the quantum emitter. This can be achieved by nanoantenna design\cite{novotny2007effective,curto2010unidirectional,feichtner2012evolutionary}.
In spatial coupling, QE perches in the plasmonic near field where closer proximity of QE to nanoantenna introduces higher coupling efficiency. 
At the single quantum system level, experimental realization of a nanoantenna's spatial coupling is a technical challenge. Nevertheless, any technique, that can challenge this successfully, has the potential to substantiate its efficacy. In this regard, normalized emission rate is the criterion in demonstrating effect of the nanoantenna\cite{novotny2012principles,karanikolas2014spontaneous}.
\begin{align}
	{\frac{\Gamma}{\Gamma_0}\equiv 
		\frac{ \rho(\vec{r}_0, \omega_0)}{ \rho_0(\vec{r}_0, \omega_0)}}
	\label{eq:1}
\end{align}
$\Gamma_0$ ($\Gamma$) is the emission rate of a QE and $\rho_0$ ($\rho$) is photonic local density of states (LDOS) at the position, $r_0$, of QE in the absence (presence) of a nanoantenna. The normalization is useful in cancelling out the transition matrix element and reducing the relation to a simple ratio in Eq.(\ref{eq:1}) which expresses ‘nanoantenna’ effect. 
It may also read as modification of LDOS on the point $r_0$ (position of the QE) at frequency $\omega_0$. The adjective \textit{local} implies that $\rho_0$ is a function of position\cite{Tigran}.
Any change in $r_0$ inherently causes a different $\rho_0$ value. That is to say, for the sake of a reliable controlled experiment and a deterministic analysis, one should not change the emitter’s position throughout the spatial coupling of a nanoantenna. \par
For example, LDOS for a QE, which rests at $r_1$, is $\rho_0(r_1, \omega)$ in the absence of a nanoantenna. Then, the QE is transported to a point, $r_2$, in the vicinity of a nanoantenna. Modified LDOS becomes $\rho(r_2, \omega)$ in which the normalized emission rate of the QE would be;
\begin{align}
	{\frac{\Gamma}{\Gamma_0} \equiv 
		\frac{ \rho(\vec{r}_2, \omega)}{ \rho_0(\vec{r}_2, \omega)}             	
}\label{eq:2}
\end{align}
and would not be, 
\begin{align}
	{\frac{\Gamma}{\Gamma_0}
		\neq   \frac{ \rho(\vec{r}_2, \omega)}{ \rho_0(\vec{r}_1, \omega)}              	
	}
	\label{eq:3}
\end{align}
The expression (\ref{eq:3}) shows that effect of nanoantenna on the QE’s emission rate is ambiguous and moving the QE is not an appropriate method for a before-and-after comparison. More precisely, in the absence of nanoantenna, the QE is considered with its environment as a whole, including the substrate. Hence, the definition of ‘deterministic coupling’ should be scrutinized thoroughly.

Deterministic coupling\cite{Benson2009} of a plasmonic nanoantenna with a single quantum emitter is first demonstrated using a scanning probe\cite{bek2008fluorescence}.
As an alternative to scanning probe technique it may be considered to fabricate nanoantenna in the vicinity of QE using conventional micro and nanofabrication techniques. In vast majority of these techniques, the locality of the QE is exposed to changes due to gas or liquid based chemical procedures often encountered in these techniques. Moreover, ensuring QE to stay intact during the procedures is not straightforward. \par
 
One suggestion is to place the QE onto a nanoantenna array\cite{Tran2017}. Although this may be preferred for the practical reasons of experimental work, it contravenes with the above discussion described by Eq.(\ref{eq:3}). Even though a value for the emission rate before and after placement can be acquired, in the strict sense, it would not be fair to suggest achievement of deterministic coupling after the transfer procedure. This is because the spatial configuration of the QE after the transfer cannot be ensured to match the configuration before the transfer during lifetime or emission rate measurements. Moreover, in order to identify a particular QE's relative position to nanoantenna, it would require performing an additional investigation with scanning near-field optical microscope.\par 	
In this work, we demonstrate a highly practical and low-cost method to obtain a quantum-plasmonic hybrid system.  It provides an opportunity to investigate luminescence features of a very same QE with and without nanoantenna in the strict sense which is consistent with the discussion around Eq.(\ref{eq:2}). The method allows one to estimate spatial configuration of the QE-nanoantenna system without having to spatially control or image the configuration using a scanning probe microscope  system. Furthermore, enhancement and quenching of radiative emission rates can be controlled through uncomplicated handling of the experimental parameters. For a demonstration of the strength of our technique, we present our results for two different settings of one of the control parameters, namely initial silver coating thickness which in-turn controls the average nanoantenna size, in our manuscript. \par
As a quantum emitter, we use defect centers dwelling in two dimensional (2D) van der Waals layered hexagonal Boron Nitride\cite{Tran2015} (hBN) multilayers. We start with hBN flakes which are multilayer quasi-2D structures that we disperse on polished silicon substrates using simple drop casting of an hBN suspension (Graphene Supermarket BN Solution). Initial recognition of an hBN flake hinges upon its silhouette (Figure \ref{fig:fig1}a, \ref{fig:fig1}b, also see Supplementing Information, Figure 3) through a coarse scanning process over a vast area on the silicon substrate using an optical microscope. The actual confirmation is made using fluorescence data collected from the candidate hBN flake under laser illumination (532 nm cw) through a spectrometer coupled to the optical microscope, a configuration which can be called a micro-photoluminescence ($\mu$-PL) setup. Exact position of a candidate hBN flake observed in optical microscope is found again in $\mu$-PL setup by using preformed unique markings on the substrate surface before deposition of hBN flakes (See Supplementing Information section of methods, Figures 1,2).  When the obtained fluorescence spectrum exhibits energy differences with zero phonon line (ZPL) peaks and its phonon side bands (PSB) as expected from hBN Raman spectrum, we use this information as a confirmation for an hBN defect center. In hunting for a defect center, excitation laser spot ($\sim$ 1.5 $\mu$m) of the $\mu$-PL setup (Figure 4 in Supporting Information) is scanned around an hBN flake till a convincing PL spectrum belonging to a defect center is captured during a real time acquisition (Figure \ref{fig:fig1}d). Fine tuning of the position is performed using a piezo nanopositioner to attain the maximum PL-signal from the color center. In addition, polarization angle-resolved $\mu$-PL capability provides polarization dependent collection of the fluorescence emission. In Figure \ref{fig:fig1}d, we present an example of the case in which a defect center has a ZPL at 705 nm clearly exhibiting dipole pattern. PSB of the emission at 780 nm  confirms that PL originates from a single hBN defect center.\par 
\begin{figure} [h!]
	\centering
	\includegraphics[scale=0.55]{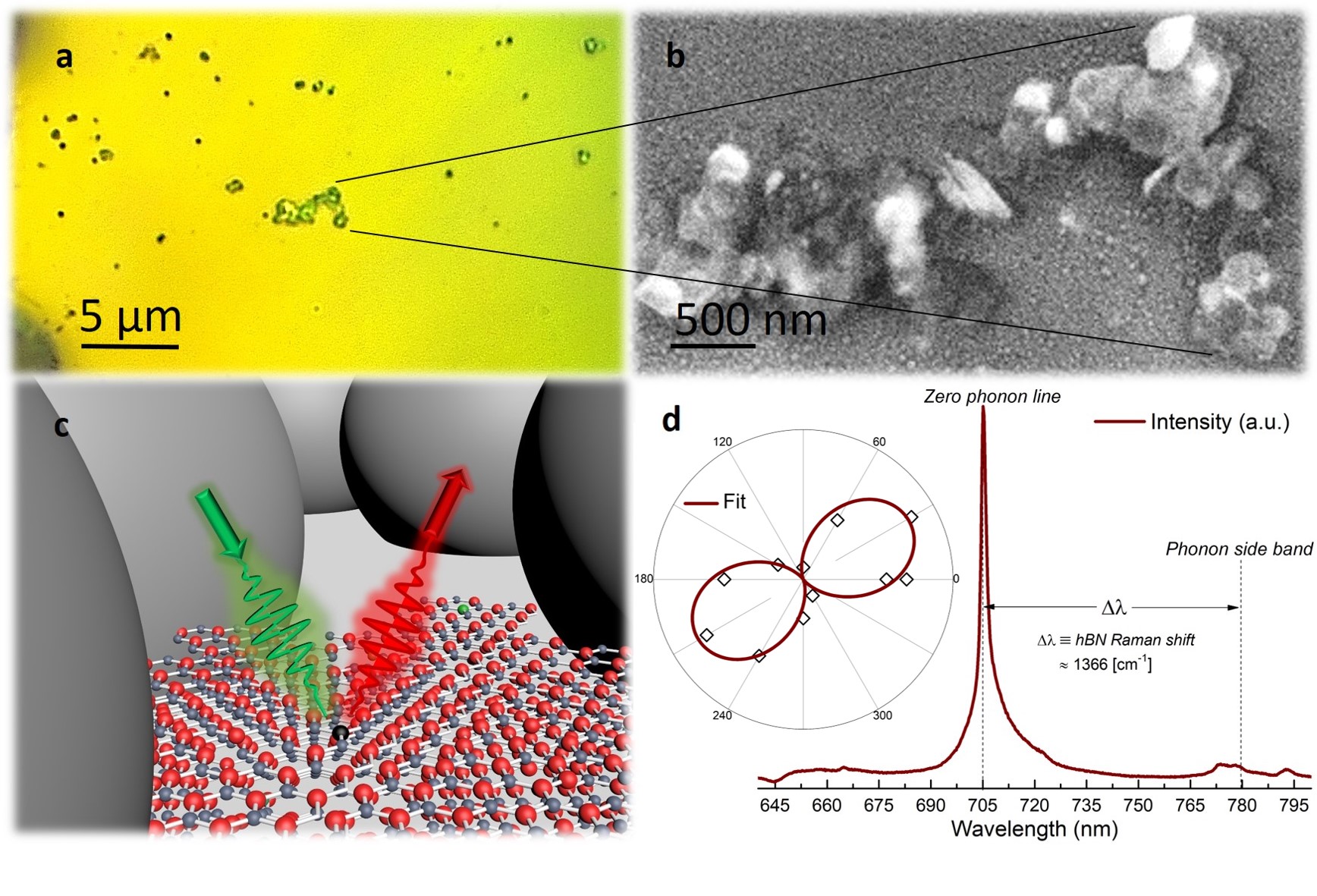}
	\caption{(a,b) Optical and scanning electron microscope images of selected hBN flakes (c) 3-D sketch of an hBN flake and Ag-nanoantennas. (d)  Typical spectrum captured from an hBN flake in which a defect center is embedded. Inset: Excitation polarization dependent ZPL intensity confirms the dipole nature of the emission. }
	\label{fig:fig1}
\end{figure}
\begin{figure} [h!]
	\centering
	\includegraphics[scale=0.48]{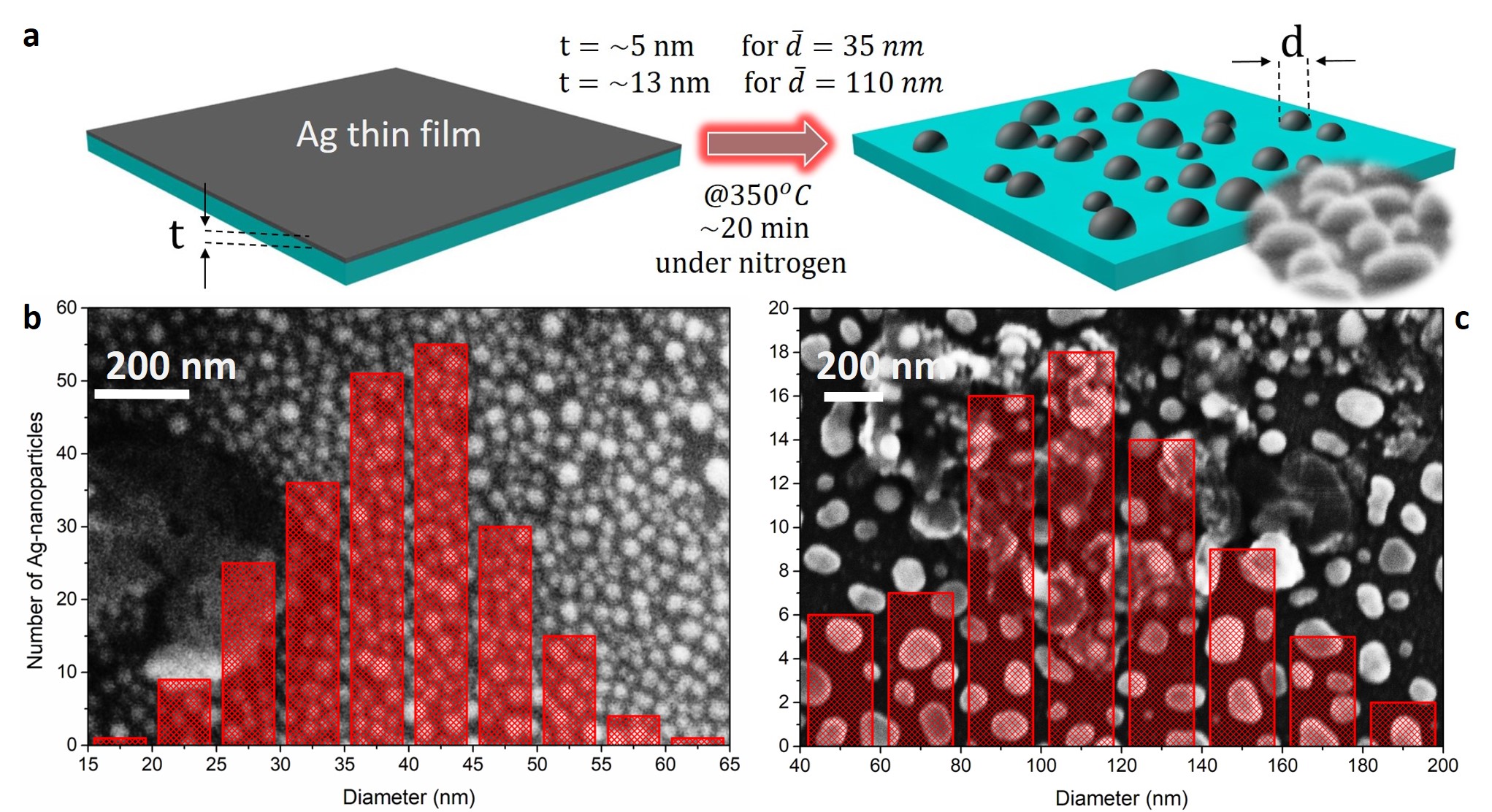}
	\caption{(a) Dewetting of an Ag thin film demonstration (b,c) Size distribution histogram and corresponding electron microscope images for two different regimes.}
	\label{fig:fig2}
\end{figure}

 
In our method, we fabricate silver (Ag) plasmonic nanoantennas in the vicinity of the defect center using a self organized approach, rather than moving the defect center to the vicinity of an already fabricated nanoantenna. We especially refrain from transferring of the defect center in order to ensure identical photonic LDOS of the environment for unambiguous determination of plasmonic nanoantenna-QE coupling. In the fabrication of silver plasmonic nanoantennas, a technique so-called `solid-state dewetting of thin films' is used in which a ~5-20 nm thick Ag film is deposited on the hBN decorated silicon substrate subsequently annealed at $\sim$350$^\circ C$\cite{Jiran1990}. Silver is selected for its low-loss, narrow and strong plasmon bands and excellent dewetting properties in comparison to other noble metals, for instance, gold. During the annealing process, high surface-area-to-volume ratio yields instability at temperatures significantly below the melting point of silver.\cite{Jiran1992} A driving force occurs to balance the instability by diffusion mass transfer. As a consequence, Ag nanoislands are formed due to the disintegration on Ag thin film. Agglomeration of nanoislands happens in varying sizes. Nanoisland size distribution is centered around a peak value where its position can be controlled using different initial film thicknesses (Figure \ref{fig:fig2}). It is remarkable that the nanoislands formed by dewetting are, to a large extent, of hemispherical or hemispheroidal shape in case of Ag, which is not always the case, for instance, in Au dewetting. Dewetting of a thin Ag film on hBN flakes creates hemispheroid Ag nanoantennas all around the hBN flakes as well as on the silicon substrate. Due to stochastic nature of self organized dewetting process, nanoislands are also formed in the vicinity of defect centers, as probably as anywhere else. The entire fabrication is just a 2-step procedure involving a vacuum thermal metal evaporation and thermal annealing in $N_2$ environment. This procedure is gentle to hBN flakes and the silicon substrate since both the defect centers and the silicon substrate are known to be very stable up to $\sim800$-$900^\circ C$ high temperatures.\cite{tran2016robust,kianinia2017robust}\par 
Electromagnetic simulations on Ag nanoparticles (AgNPs) show that the particle size and quality factor ($Q$) are inversely proportional in literature\cite{koenderink2010use} confirmed by our own simulations (see Supporting Information section on simulations). Thus, an AgNP size at $\sim$30 nm enables supporting a $Q$ of $\sim$10-20, which is considerably high when compared to that of a larger AgNP of for example $\sim$100 nm size with $Q$<3. However, a 30 nm AgNP resides in absorption dominant regime\cite{Absorption} while a 100 nm AgNP resides in scattering dominant regime\cite{Scattering} in the extinction efficiency. For this reason, we present our results for two distinctly different regimes; case-1 and case-2 which correspond to 35  nm (Figure \ref{fig:fig2}b) and 120 nm (Figure \ref{fig:fig2}c) average size Ag plasmonic nanoantenna, respectively.\par 
\begin{figure} [h!]
	\centering
	\includegraphics[scale=0.64]{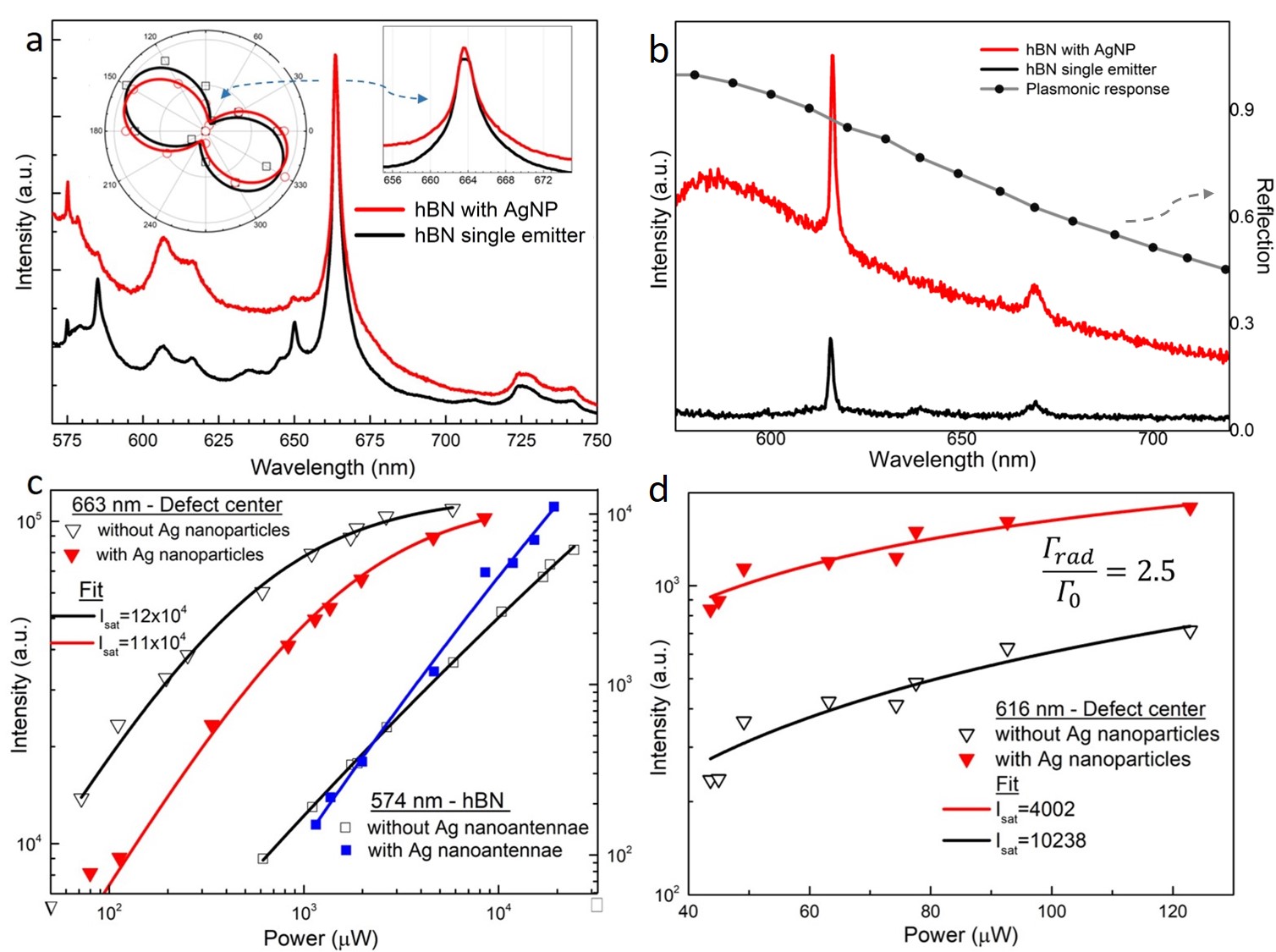}
	\caption{(a) PL Spectra for the defect center at 663 nm. The black and red lines are for before and after small size nanoantennas. The inset shows polarization dependency of the defect at 663 nm.  (b) Spectra for the defect center at 616 nm. The red line for after large size nanoantennas. (c,d) Power-resolved spectra in log-log scale (see Supplementing Info section on data analysis, Figure 5-9).}
	\label{fig:fig3}
\end{figure}
Selected defect centers have ZPL at $\lambda=663$ nm for case-1 and $\lambda=616$ nm for case-2 (Figure \ref{fig:fig3}) whereas corresponding PSBs at 729 nm and 672 nm are the signs of hBN host lattice with 1366 $cm^{-1}$ Raman shift due to its $E_{2g}$ in-plane phonon mode.\cite{Tran2015}
Ag nanoislands are fabricated on hBN flakes. Figure \ref{fig:fig3}a and \ref{fig:fig3}b  show the effect of Ag plasmonic nanoantennas on the emission spectra for $\lambda$=663 nm and  $\lambda$=616 nm, respectively. In each plot, the PL of a particular single defect center with (in red) and without (in black) the Ag plasmonic nanoantennas are given. Defect centers are irradiated under the same conditions. Since the measurement of plasmonic response on hBN flake is not possible, we performed a reflection measurement with similar size AgNPs on Si/Indium Tin Oxide (ITO) substrate which has same refractive index (n$\simeq$1.8) as that of Si/hBN in visible spectrum \cite{hBNindex} (Figure \ref{fig:fig3}b). Power dependent $\mu$-PL measurements for two cases before and after Ag nanoantennas fabrication are given in Figure \ref{fig:fig3}c and \ref{fig:fig3}d. For case-1, Ag nanoantenna size of ~35 nm being in the absorption dominant regime causes quenching of the defect emission which is evident from the reduction in the background subtracted ZPL intensity. At the same time, the broadband background due to Raman scattering of hBN lattice appears to have enhanced which is eminent of expected surface enhanced Raman scattering (SERS) effect due to dense decoration of the flake with  Ag plasmonic nanoantennas.\cite{fleischmann1974raman, jeanmaire1977surface, kneipp1997single}. For case-2, a self evident enhancement is observed in Figure \ref{fig:fig3}d with the scattering dominant regime of ~110 nm sized Ag nanoantennas.\par  
In the framework of nanoantenna picture, incoming electromagnetic field of excitation source is captured by the nanoantenna based on its interception area (extinction cross section). Depending on the particle size, captured electromagnetic field is either predominantly absorbed (Joule loss) or is scattered. Substantial amount of electromagnetic energy is coupled to the surface plasmons of nanoantenna and is localized in its near field (often referred to as hotspot formation). Depending on the nature of the coupled surface plasmon mode being dipole or higher order-like, the detectivity of the scattered photons in the far field varies. In the framework of a defect center in the vicinity of a nanoantenna there are two physical processes responsible for the fluorescent enhancement and lifetime reduction. Firstly, both the direct incident photons and the nanoantenna scattered photons excite the defect center (referred to as field enhancement) which is responsible for enhanced fluorescence without modification of the lifetime. Secondly, a part of the fluorescent photons upon scattering from the nanoantenna will be directed back to the defect center that result in stimulated emission of photons with consequence both in lifetime shortening and fluorescence enhancement. Subsequently, the defect center emits more photon at a time. This corresponds to the Purcell factor\cite{Purcell} ($\tau_0/\tau$) associated with the effect of plasmonic nanoantenna on the emission properties of the QE, in which is denoted by $\tau_0$ and $\tau$ are the lifetimes in the absence and presence of nanoantenna, respectively. Note that detected photons in the far field do not alone stand for the Purcell factor since a significant part of photons that are localized in the near field of the nanoantenna\cite{Hoang2015}. Due to the fact that depending on the size of the nanoantenna, a significant portion of the localized photons (in the form of surface plasmon polaritons- SPPs) are subject to Joule losses (or absorption in the metal). So it can be stated that a significant part of the emission can be of nonradiative nature due to absorptive losses. Since the curves provided in Figure \ref{fig:fig3}d correspond to far-field detected photons, the measured enhancement factor of $\sim$2.5 corresponds to normalized "radiative" emission ($ \Gamma_{rad} / \Gamma_0 $). The effect can be safely extended to a discussion on the modification of quantum yield with and without the presence of the Ag nanoantennas, as often the native quantum yield of QEs tend to suffer from intrinsic nonradiative losses.\cite{Mohammadi2008}  \par
\begin{figure} [h!]
	\centering
	\includegraphics[scale=0.62]{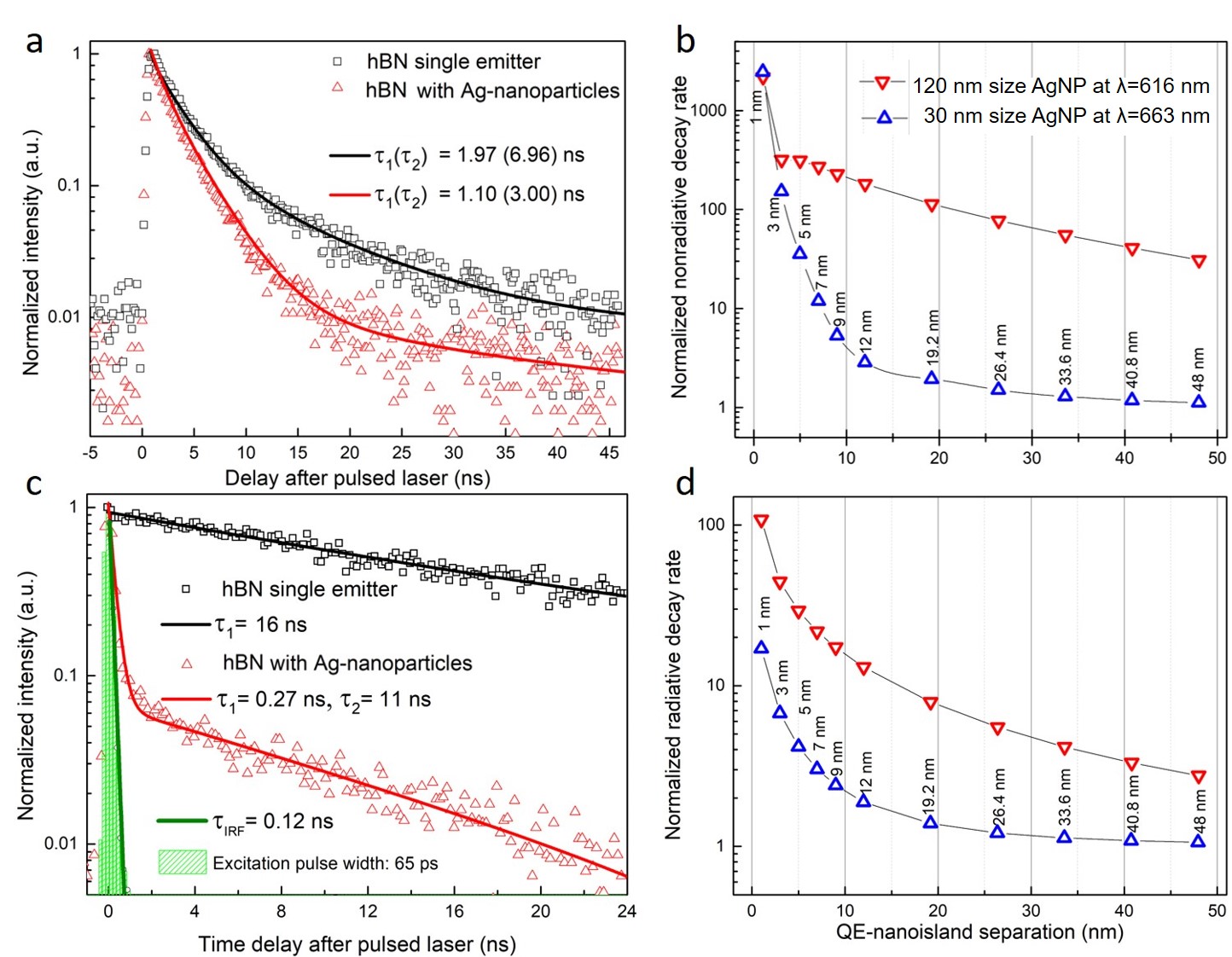}
	\caption{(a) Time-resolved measurements for the defect center at 616 nm with large size nanoantennas (b) Simulation results of normalized nonradiative decay rate for case 1 (blue triangles) and case 2 with two simultaneous nanoantennas (red triangles) in the vicinity of hBN defect center. (c) Time-resolved measurements for the defect center at 663 nm with small size nanoantennas. (d) Simulation results of normalized radiative decay rate values for for case 1 (blue triangles) and case 2 (red triangles) with two simultaneous nanoantennas (red triangles) in the vicinity of hBN defect center.}
	\label{fig:fig4}
\end{figure}
\begin{align}
	\textit{Purcell Factor}={\frac{\Gamma}{\Gamma_0}=\frac{\tau_0}{\tau}= \frac{\Gamma_{rad}+\Gamma_{nonrad}}{\Gamma_0}          	
	}
	\label{eq:4}
\end{align}
Time-resolved investigation of hBN defect centers with and without nanoantenna clearly demonstrates lifetime reduction in both defect centers (Figure \ref{fig:fig4}). Purcell factor can be both expressed as the ratio of the QE transition rate with the nanoantenna to native QE transition rate, or as the inverse ratio of the lifetime of the QE with the nanoantenna to native QE lifetime\cite{Purcell}. In case of nanoantenna, the transition rate is the sum of radiative and nonradiative transition rates. In Figure \ref{fig:fig4}a, we show the time resolved fluorescence measurements on an hBN defect center in the absence and presence of a Ag plasmonic nanoantenna in case 1. Our analysis shows that the time response can be modeled by a double exponential decay function in both cases. The two initial  time constants of 1.97 and 6.96 ns determined in the absence of the nanoantenna are found to reduce to 1.10 and 3.00 ns, respectively. In Figure \ref{fig:fig4}a, Purcell factor is determined to be $\simeq$ 2 for the defect center in case-1. We have conducted electromagnetic simulations using finite elements method (FEM) in order to determine normalized radiative decay rates as a function of Ag plasmonic nanoantenna distance to the hBN defect center. Since the Purcell factor is due to collective influence of normalized radiative and nonradiative rates, the simulation results given in Figure \ref{fig:fig4}b and \ref{fig:fig4}d enable to estimate a spatial configuration of defect-nanoantenna system. For instance, a 35 nm Ag hemispheric nanoantenna placed at 40-48 nm distance from the defect center provides for suitable conditions as in case-1. It should be noted that, size deviations from 35 nm can give rise to change of the radiative and nonradiative contributions to the sum. This can be verified experimentally. \par 
The spatial configuration of case-2 can be estimated in a similar way. Figure \ref{fig:fig4}c
shows time-resolved fluorescence measurements on an hBN defect center revealing an initial time constant of 16 ns dropping down to two time constants of 0.27 and 11 ns which correspond to Purcell factors of 60 and 1.5, respectively. Note that all of the measured time constants are significantly larger than the instrumental response function time scale of $\tau_{IRF}$=0.12 ns (green curve). Regarding the simulations, red triangles in Figure \ref{fig:fig4}b, \ref{fig:fig4}d represent the condition for a 120 nm size Ag-hemisphere calculated at $\lambda$=616 nm. On the one hand, an estimate for the distance of the Ag plasmonic nanoantenna to the defect center yields 30 nm considering that Purcell factor contributions from normalized nonradiative rate is 55 and normalized radiative rate is 5 at this distance. On the other hand, one does not have to estimate the radiative contribution to the Purcell factor because it is already experimentally determined to be ($\Gamma_{rad}/\Gamma_0$=2.5) as in Figure \ref{fig:fig3}d. That readily sets the nonradiative contribution to be at a value of  ($\Gamma_{nonrad}$/$\Gamma_0$=57.5) in order to yield a Purcell factor 60. The simulation results provided in Figure \ref{fig:fig4}d suggests the defect center to Ag plasmonic nanoantenna separation to be at 45 nm for a radiative contribution of 2.5 but the nonradiative part suggests a 35 nm separation. In light of this, one may realize that two explicitly different decay times in Figure \ref{fig:fig4}c strongly indicate the possibility of two simultaneous nanoantennas playing a role in case 2. Therefore, in this regard we have performed a simulation accounting for two simultaneously interacting nanoantennas with the hBN defect center.\par 
\begin{figure} [h!]
	\centering
	\includegraphics[scale=0.6]{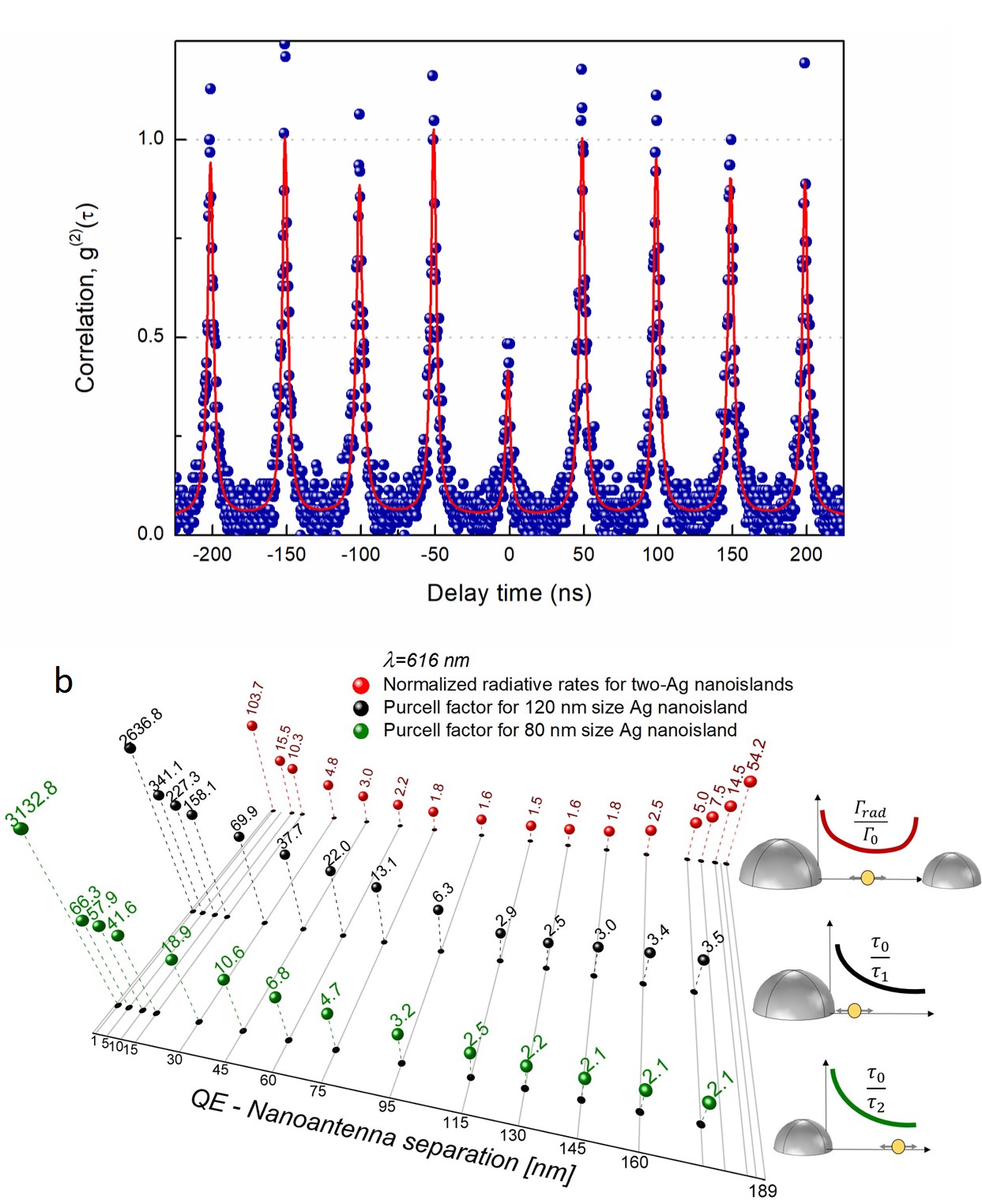}
	\caption{For scattering dominant regime (case-2); a) Photon correlation measurement of hBN defect center-AgNP system and b) Purcell factor and normalized radiative rate values for different positions of hBN-defect center between 80 nm and 120 nm size Ag nanoantennas (see Supplementing Information on simulations, Figs. 10-25).}
	\label{fig:fig5}
\end{figure}
We have used a Hanbury-Brown-Twiss (HBT) interferometer for characterization of emission properties of QEs (details of the setup is given in Methods). Figure 5a shows a second-order correlation function, $g^{(2)} (\tau)$, measured from the emitter coupled with Ag plasmonic nanoantenna in case 2. The dip with a value below 0.5 shows at zero delay time (without background correction) confirms that the emitter is indeed a single photon source.
Figure \ref{fig:fig5}b depicts the case in which two Ag hemispherical nanoantennas 80 nm and 120 nm size are placed at a 190 nm distance to each other. The position of hBN defect center is scanned between the two nanoantennas along the axis connecting their centers. The radiative rates are computed and normalized to corresponding values that are computed in the absence of nanoantennas. In addition, each Ag hemispherical nanoantenna is also considered separately in order to calculate their individual contribution to the Purcell factor. For case-2, based on experimental values of radiative rates in Figure \ref{fig:fig3}d and Purcell factors in Figure \ref{fig:fig4}c, a value can be appointed to the Ag nanoantenna-hBN defect center distance following Figure \ref{fig:fig5}b. To be more clear, if one follows the red dots in Figure 5b, it appears that a Purcell factor of 2.5 is achieved at the QE-Nanoantenna separation of the 160 nm. This means that the QE resides at a position which is at a 160 nm distance to one nanoantenna and at a 30 nm distance to the other one. Purcell factors of 60 and 2 are obtained at 32 nm and 160 nm distances following the black and green dots in Figure \ref{fig:fig5}b. The simulation results suggest that two simultaneous Ag nanoantennas with different sizes and distances affecting the defect center's fluorescence lifetime can be a reasonable explanation of observed two decay times in Figure \ref{fig:fig4}c. In summary, we demonstrate that dewetting based nanoantenna fabrication technique when accompanied with computer simulations can be an effective methodology for performing deterministic coupling experiments on quantum-plasmonic hybrid systems.\par

\begin{acknowledgement}

We acknowledge financial support from TUBITAK project no: 118F119. The author thanks Selcuk Yerci for the work station, at which we performed our computations, Vasilis Karanikolas for reviewing some of the calculations.

\end{acknowledgement}

\begin{suppinfo}
\begin{itemize}
  \item Supporting Information: Experimental setup and methods, substrate preparation, additional simulation results, data analysis, fit functions.
\end{itemize}
\section{Methods}
First, a 10 mm x 10 mm polished Si wafer chip is laser engraved with unique individual marks (such as Arabic numerals) in a 2D layout for finding the same particular locations under the microscope before and after processes. Then, a thorough cleaning process is made after laser marking. \par \noindent
\begin{figure} [h!]
	\centering
	\includegraphics[scale=0.5]{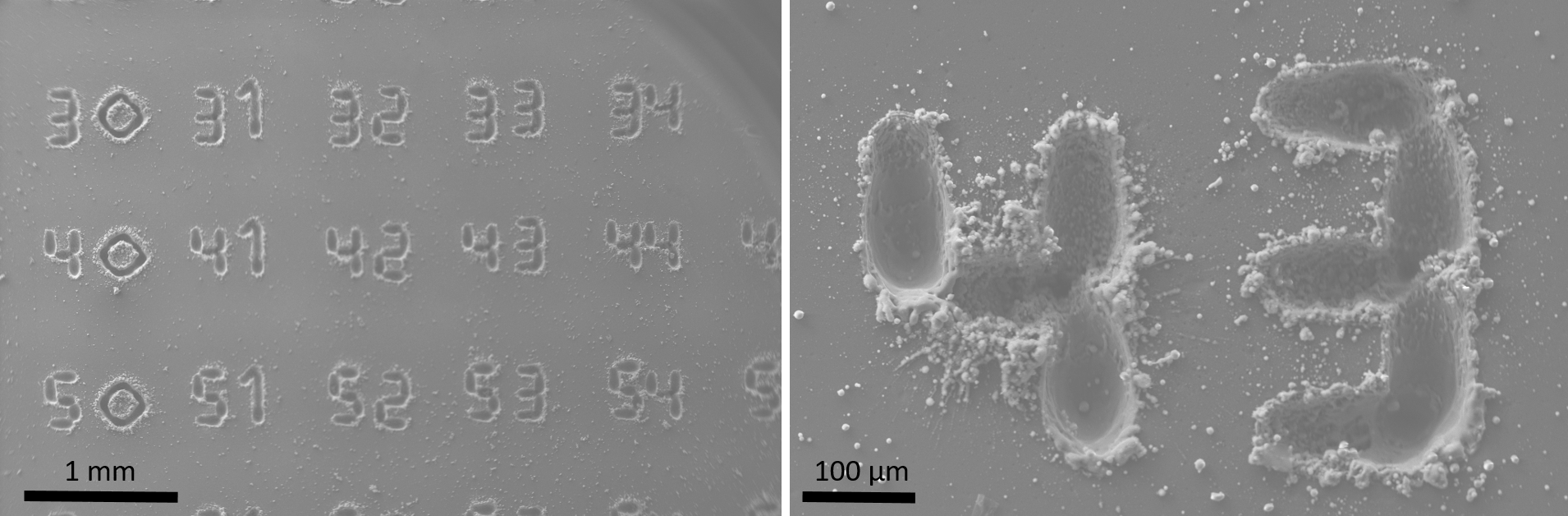}
	\caption{SEM image of laser engraved Si substrate.}
	\label{fig:fig1}
\end{figure} 
\begin{figure} [h!]
	\centering
	\includegraphics[scale=0.5]{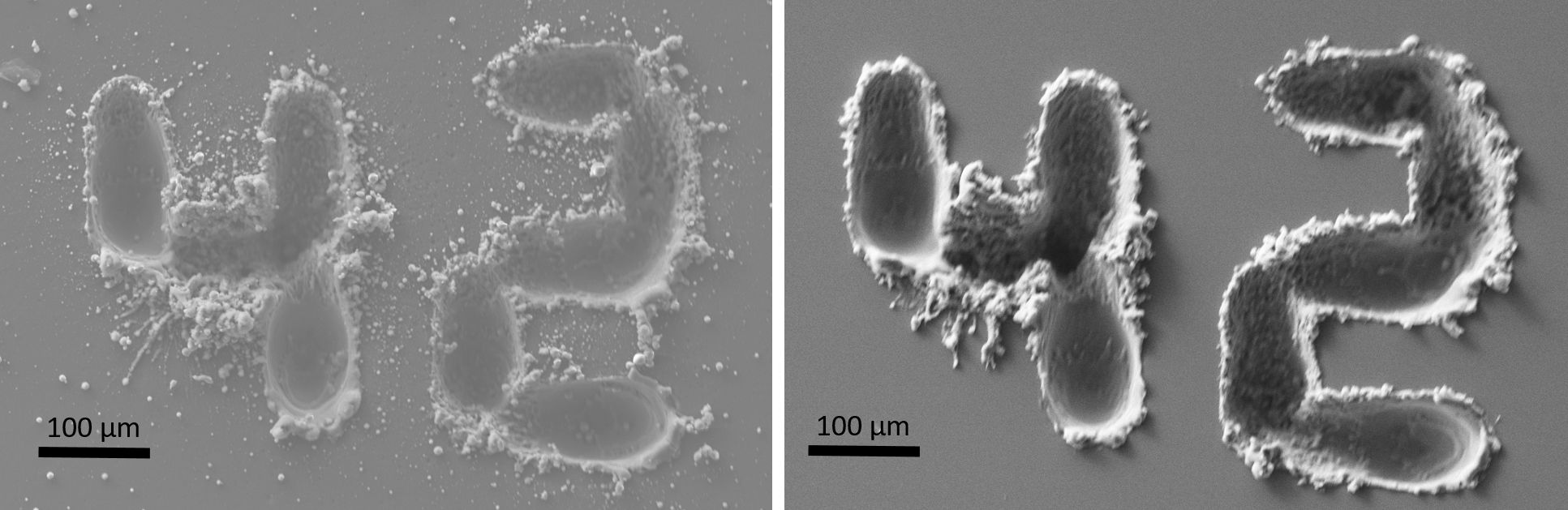}
	\caption{SEM image of same position before and after standart clean 1-2 (RCA).}
	\label{fig:fig1}
\end{figure} \\
hBN Flakes (Graphene Supermarket) are drop cast onto the substrate and annealed at $350^\circ C$ for $20$ $min$. 
The purpose of the annealing process is to assure that the selected defect is not affected in that temperature level later on during annealing step for dewetting although the fact that hBN defect centers are durable to much higher temperature annealing.
\begin{figure} [h!]
	\centering
	\includegraphics[scale=0.5]{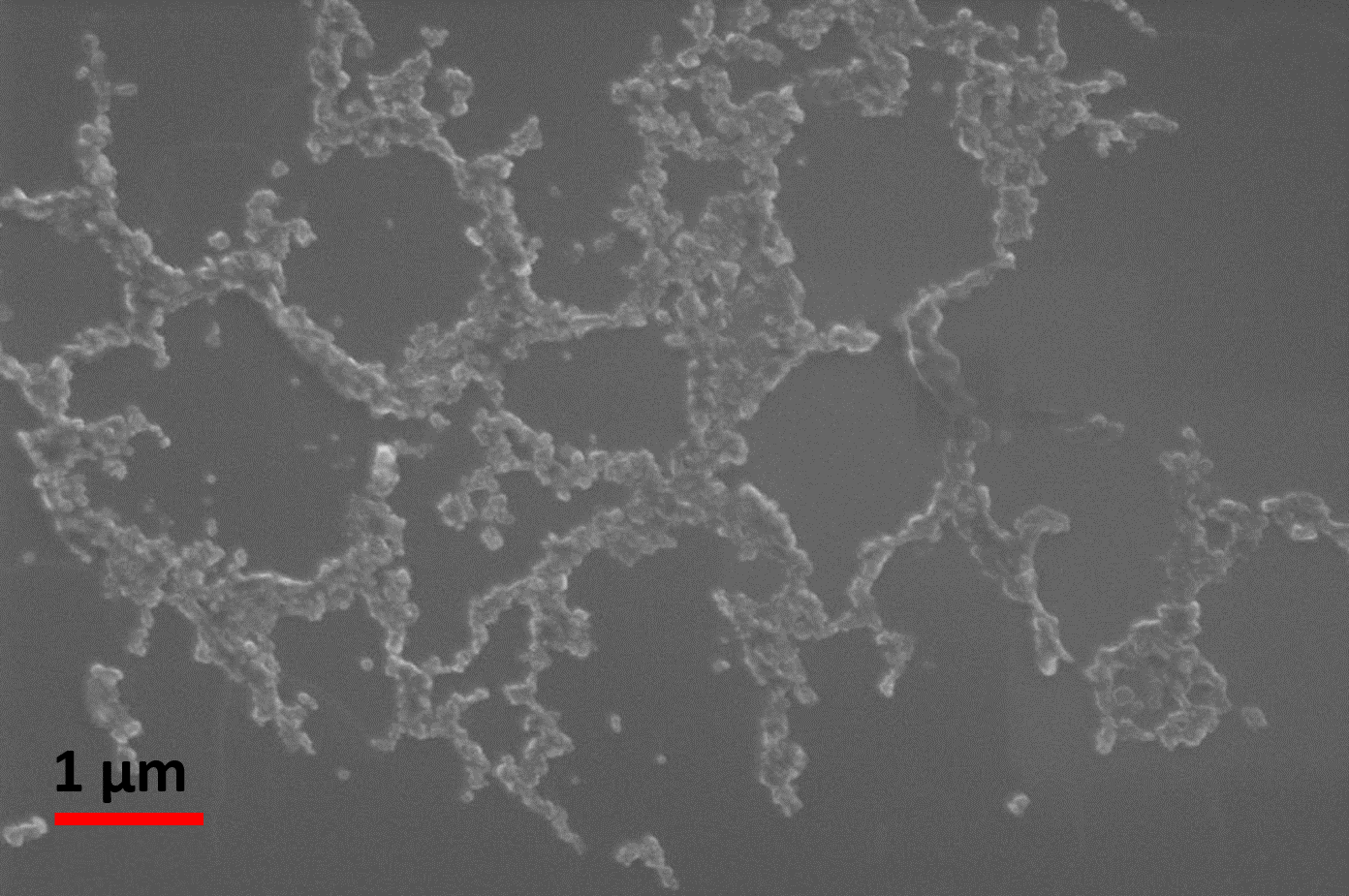}
	\caption{SEM image of a typical hBN flake.}
	\label{fig:fig1}
\end{figure} \\
\noindent
The defect hunting in an hBN flake is performed using an xyz nanopositioner stage which is assembled into a micro-photoluminescence ($\mu$-PL) setup and a Hanbury-Brown-Twiss (HBT) interferometer. 
Immediately after all the measurements of the identifed defect are done, the chip is moved into a vacuum chamber for physical vapor deposition of thin Ag film. High purity Ag source is thermally evaporated onto the chip to a thickness of 5-6 nm and 12-13 nm for two cases. 
The sample is heated at $350^\circ C$ for $20$ $min$ under nitrogen flow. The dewetted silver islands form hemispheroid particles in different sizes depending on the film thickness.
hBN defect centers are excited by a cw laser ($\lambda_{exc}$=532 $nm$, Verdi-V6 Coherent) and a pulsed diode laser ($\lambda_{exc}$=483 $nm$, 65 $ps$ pulse width, Advanced Laser Diode Systems). Fluorescence is collected using a 50X/0.75 NA objective (Optika). The optimum polarization angle of the excitation is determined using a motorized half-waveplate (HWP). 
In the detection part of the setup, the Rayleigh scattered light is rejected with a $540$ nm  notch filter. The sample surface is imaged by a CMOS camera white light illumination. A spectrometer (Andor Shamrock 750) is employed with a 3 MHz EMCCD camera (Andor Newton).
The lifetime measurement of the quantum emitter is performed by a photon counting avalanche photodiode (APD) mounted on the HBT interferometer. APDs are connected to a time tagging electronic module (TTM8000, Roithner Laser Technik).
\begin{figure} [h!]
	\centering
	\includegraphics[scale=0.46]{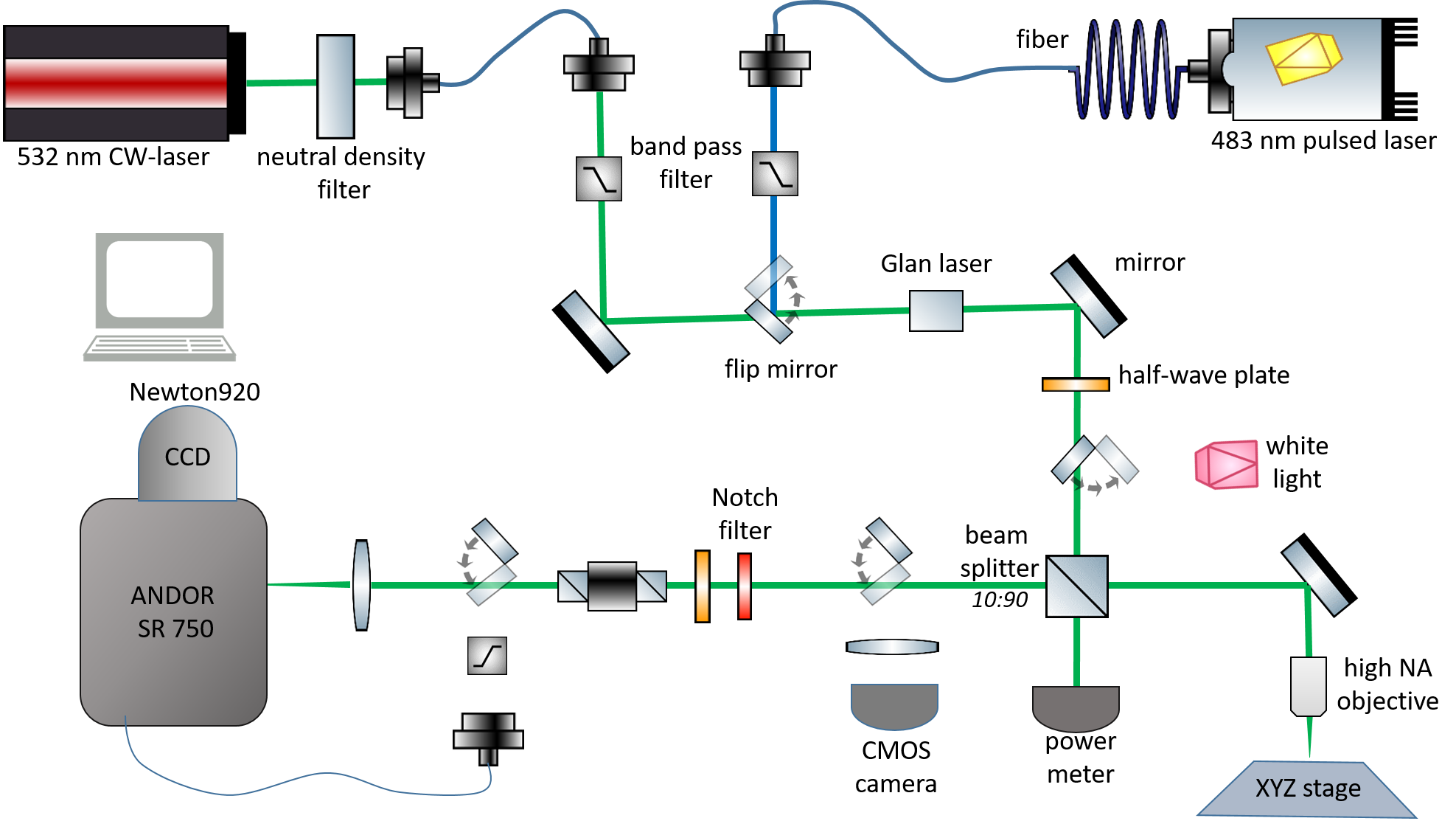}
	\caption{Experiment setup.}
	\label{fig:fig1}
\end{figure} 
\newpage
\section{Data analysis}

\begin{figure} [h!]
	\centering
	\includegraphics[scale=0.5]{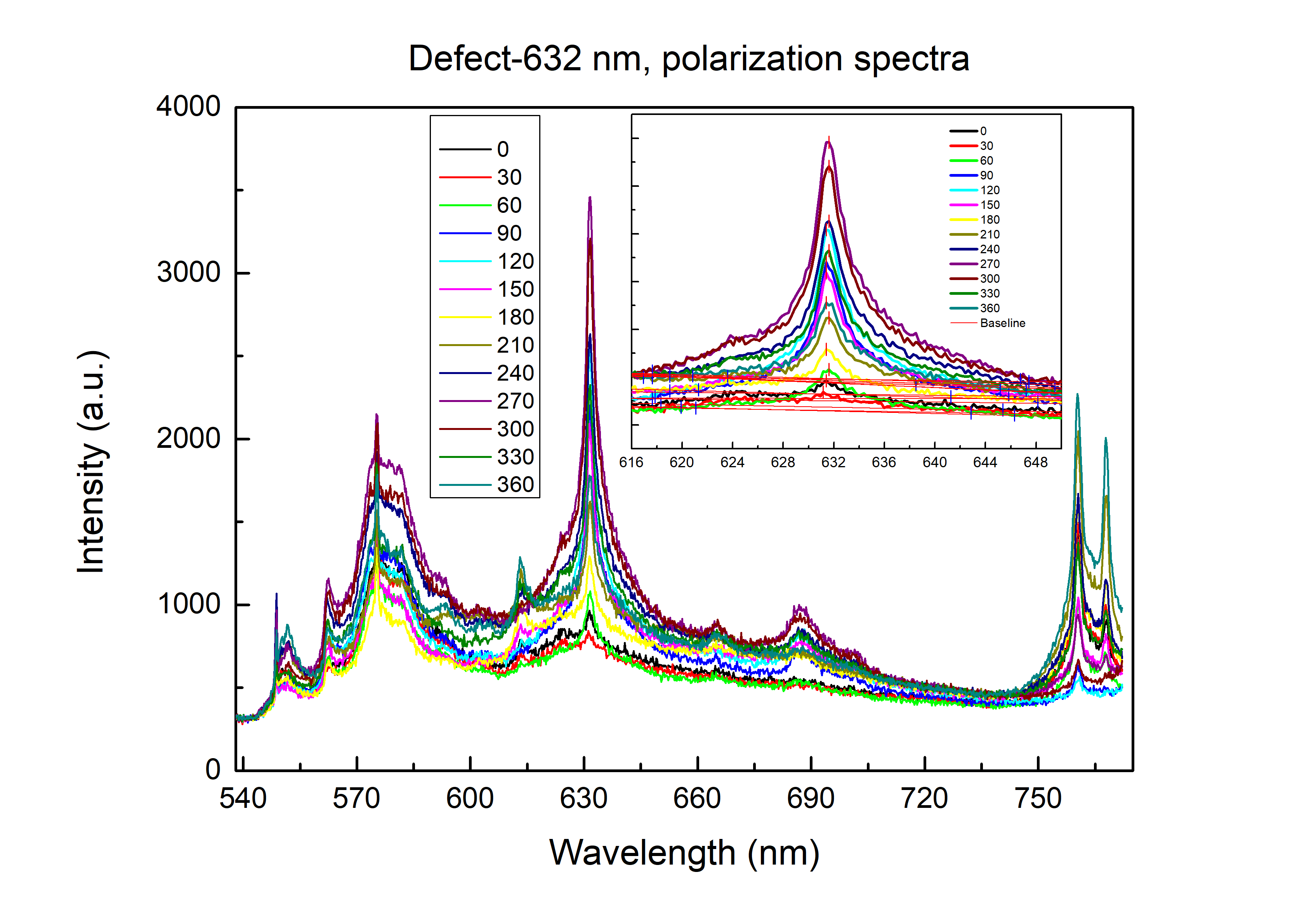}
	\caption{Polarization-resolved measurement of defect-632nm. Inset: Spectra between 616-650 nm with baselines}
	\label{fig:defect532_pol}
\end{figure} 
\begin{figure} [h!]
	\centering
	\includegraphics[scale=0.46]{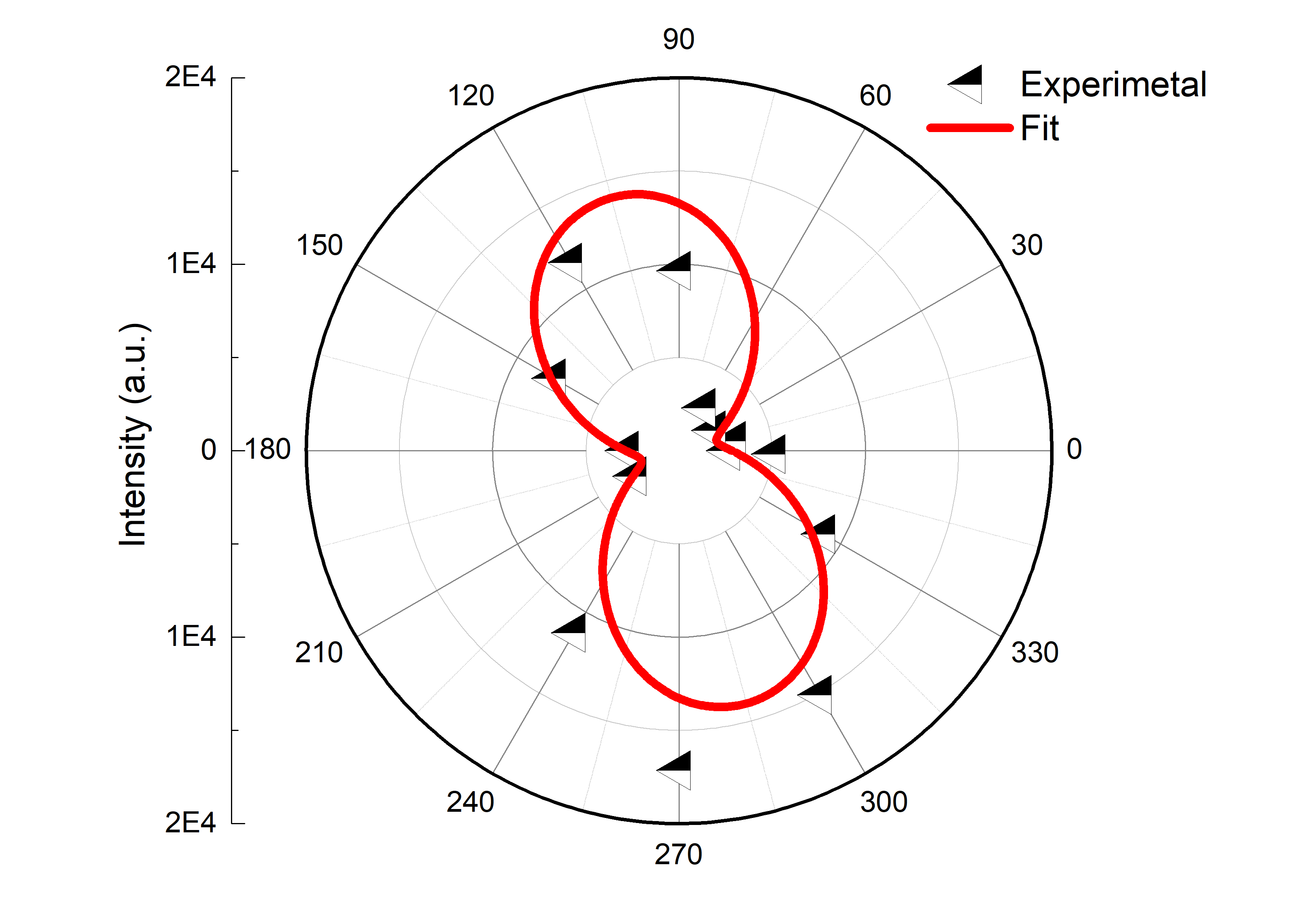}
	\caption{Polar plot of the selected data and fit}
	\label{fig:defect532_pol_}
\end{figure}
The photoluminescence intensity is captured while polarization angle changes. Since there are background effects in the spectra, each of them are substracted. 
Using the peak analyzer of OriginPro, areas below baselines are substracted (inset of figure \ref{fig:defect532_pol}). Data sets are plotted as a function of corresponding angles (figure \ref{fig:defect532_pol_}). 
\newpage
\par
Fitting function of the polar plot is chosen as $cos^2\theta$ \cite{Benson2009, Tran2015}. The function is given as;
\begin{align*}
y=y_0+A cos^2(\theta \frac{\pi}{180}+x)
\end{align*}
\begin{align*}
y&= \textit{Dependent variable}, \\
\theta&= \textit{Independent variable}, \\
y_0,A,x&=\textit{Parameters}
\end{align*} 
\newpage 
\begin{figure} [h!]
	\centering
	\includegraphics[scale=0.46]{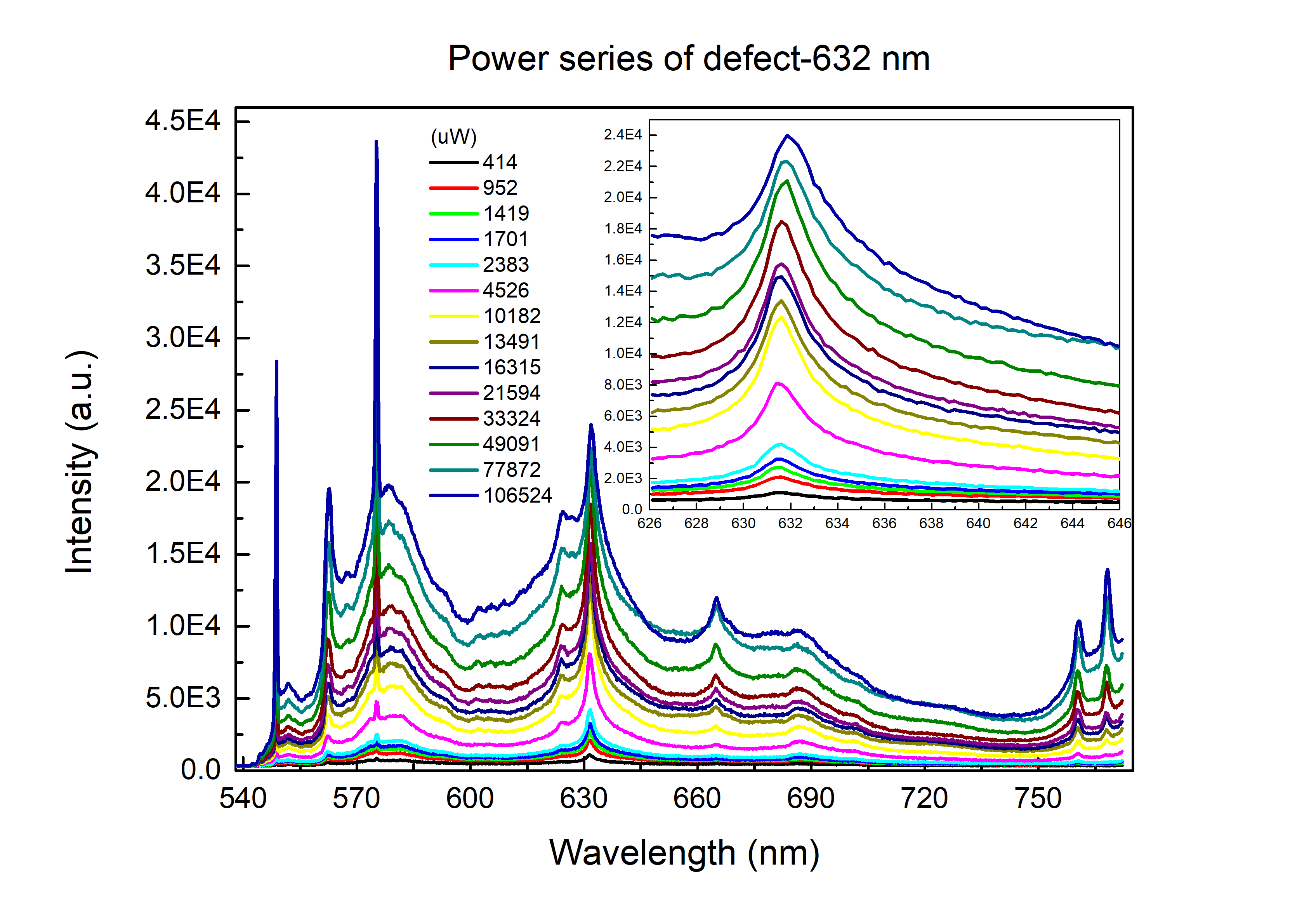}
	\caption{Power measurements of the defect-632nm. Inset: Spectra between 626-646 nm.}
	\label{fig:defect532_pow}
\end{figure} 
\begin{figure} [h!]
	\centering
	\includegraphics[scale=0.46]{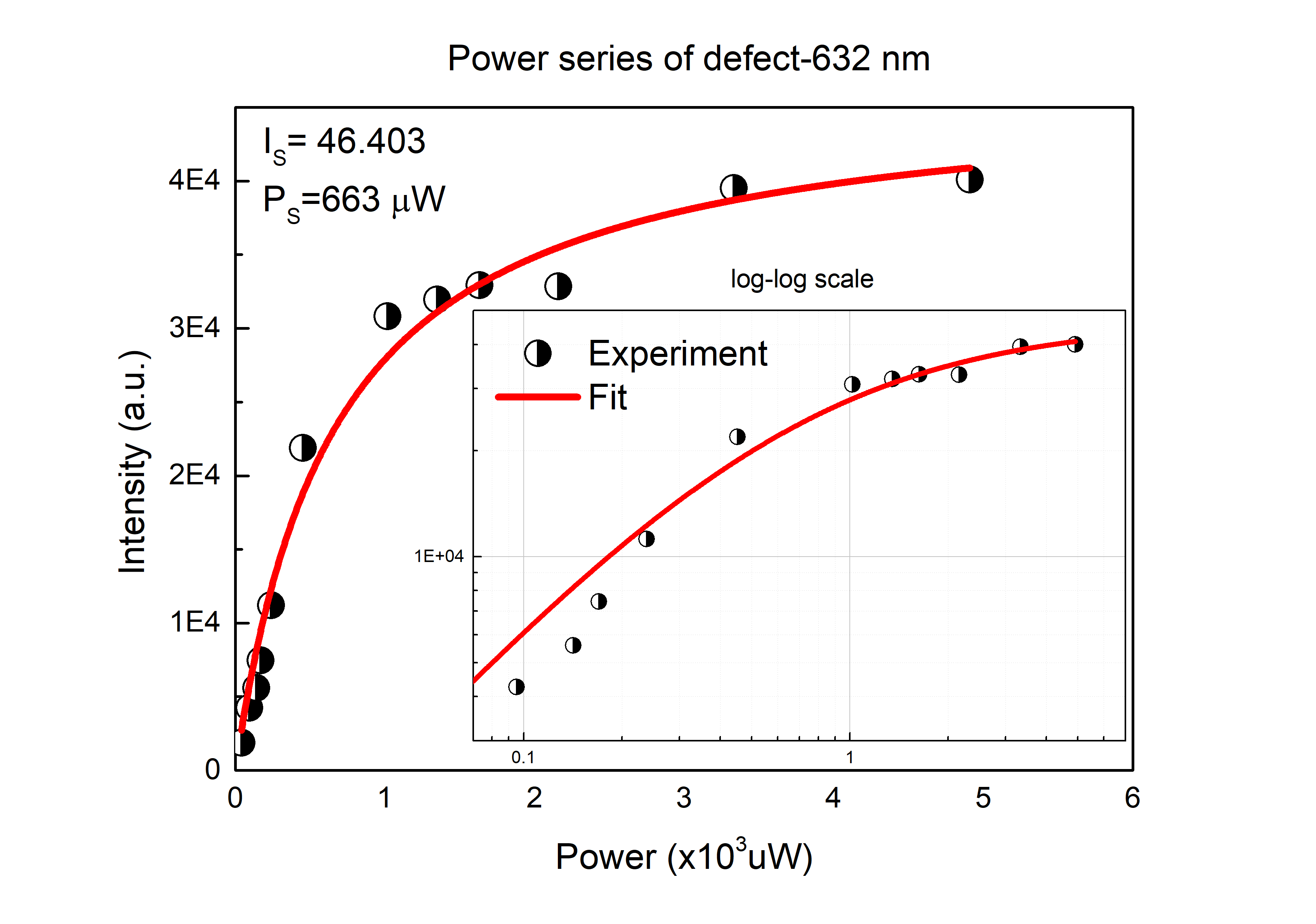}
	\caption{b) Power saturation plot and fitting of the experimental data. Inset: same plot with log-log scale.}
	\label{fig:defect532_pow_}
\end{figure}
Output power of the laser is attenuated with a round shape neutral density filter in order to increase the power gradually. The emission spectra is captured for different excitation power in figure \ref{fig:defect532_pow}. Same procedure above is followed for background substraction. Area under the intensity plot gives the power values. In a selected wavelength interval (see inset of \ref{fig:defect532_pow}), the data analysis is performed and results are plotted in figure \ref{fig:defect532_pow_}. The saturation power is observed as 663 $\mu W$. The fitting function is given as \cite{Tran2015};
\begin{align*}
I=I_{\infty} \times \frac{P}{P+P_{sat}}
\end{align*}
\par 
\newpage  
\begin{figure} [h!]
	\centering
	\includegraphics[scale=0.4]{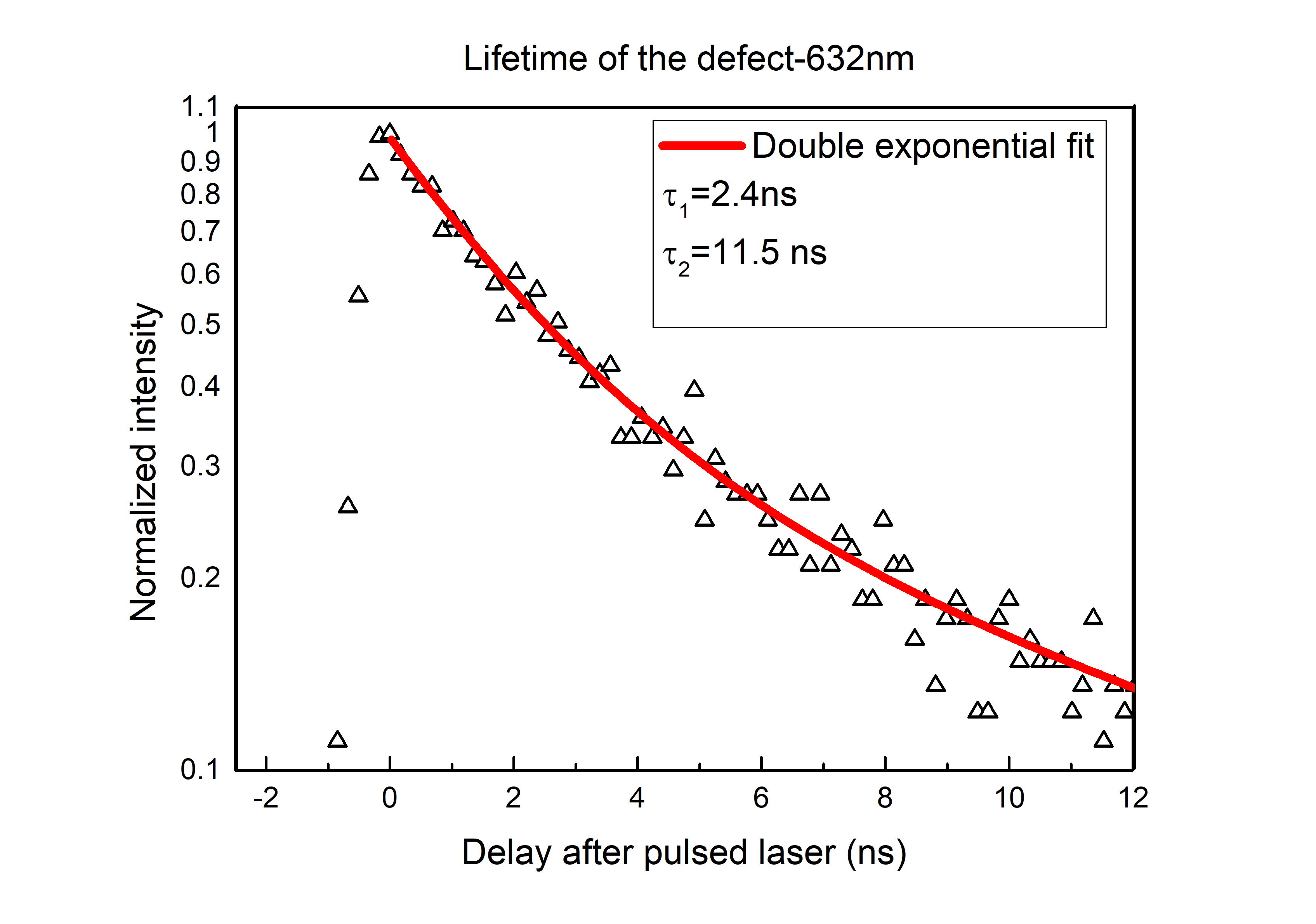}
	\caption{Semilog plot of the fluorescence lifetime with a double-exponential fit.}
	\label{fig:time_fit}
\end{figure}
Fluorescence lifetime measurement is very sensitive technique. Photons from the excitation source and the quantum emitter are correlated with a time tagging module. Basically, just after the short-pulse excitation, arrival times of photons, which are radiated from the emitter, are measured. The arrival time of each photon is tagged by a time-correlator. The difference between two time measurements shows the delay after excitation\cite{masters2008principles}. Laser pulse starts and the first photon emitted from the defect stops the counting by triggering the detector. 

The results are captured as it is seen in figure \ref{fig:time_fit}. In order to make a  data analysis, one measurement cycle is selected. The exponential decay in figure \ref{fig:time_fit} is fitted with the function;
\begin{align*}
y=y_0 + A_1 exp{(-x/\tau_{1})} + A_2 exp{(-x/\tau_{2})} 
\end{align*}
where $A_1$ and $A_2$ are fitting parameters. Since we have three-level model, double-exponential is used as a fitting function. Thereby, lifetime of the excited state, $\tau_{1}$, is found as 2.4 ns while the metastable state lifetime, $\tau_{2}$, is 11.5 ns. 
This is the expected result. One comment about fitting may be that playing with the fitting function or fitting parameters can easily cause a small changes in lifetimes. This is ignorable. Especially, within this work, the comparison of lifetime is important. As long as fitting functions are same for different experiments, lifetimes can be compared safely.   
\newpage
\section{Simulation}
First of all, an incoming plane wave is traversed through an air domain surrounded by perfectly match layers (PML). Then, a single silver nano-sphere is planted into the air domain. Later on, the substrate is introduced in the simulation domain. 
The energy attenuated in the incoming light (the extinction) and the energy localized near to the particle surface are investigated for different geometries. As a result, four different field components: Incoming fields, scattered fields, localized fields and total fields are evaluated. Optical responses of silver and silicon substrate are taken into account by the interpolation list that is defined experimentally \cite{Johnson1972}.\par 
A 25 nm radius of silver (Johnson \& Christy \cite{Johnson1972}) nanosphere is planted into the model. Reliability of computed results are tested by literature values.\cite{Bohren1998, Dereux2012} For 25 nm radius silver particle results are used for reliability check. The resonance wavelength of the particle is calculated as $~$364 nm. As it is seen in the figure (\ref{Fig:2.4}) , the quality factor is found as 24. The Drude model calculation for the same scenario in \cite{Dereux2012} including bound electron contributions show that  the quality factor is 24 however it gives the value of 21 when it is using tabulated data (Johnson \& Christy). Also, it is observed that the resonance frequency does not match exactly. Contrary to this, the \href{http://nordlander.rice.edu/miewidget}{Bohren \& Huffman code}\cite{Bohren1998} using Johnson \& Christy parameters confirms our model so that the resonance wavelength and the quality factor values are exactly same. One may want to check this code by him/herself can visit the following link which is available as an online widget \href{http://nordlander.rice.edu/miewidget}{(http://nordlander.rice.edu/miewidget)}. Moreover, the small shifts are observed for the same case of scattering on silver sphere in different approaches. Analytical methods and their numerical evaluations are clearly giving different spectral values than calculations which are made with 'full electrodynamics solution'.\par
\begin{figure}[!htb]
	\begin{minipage}{0.5\textwidth}
		\centering
		\includegraphics[width=1.1\linewidth]{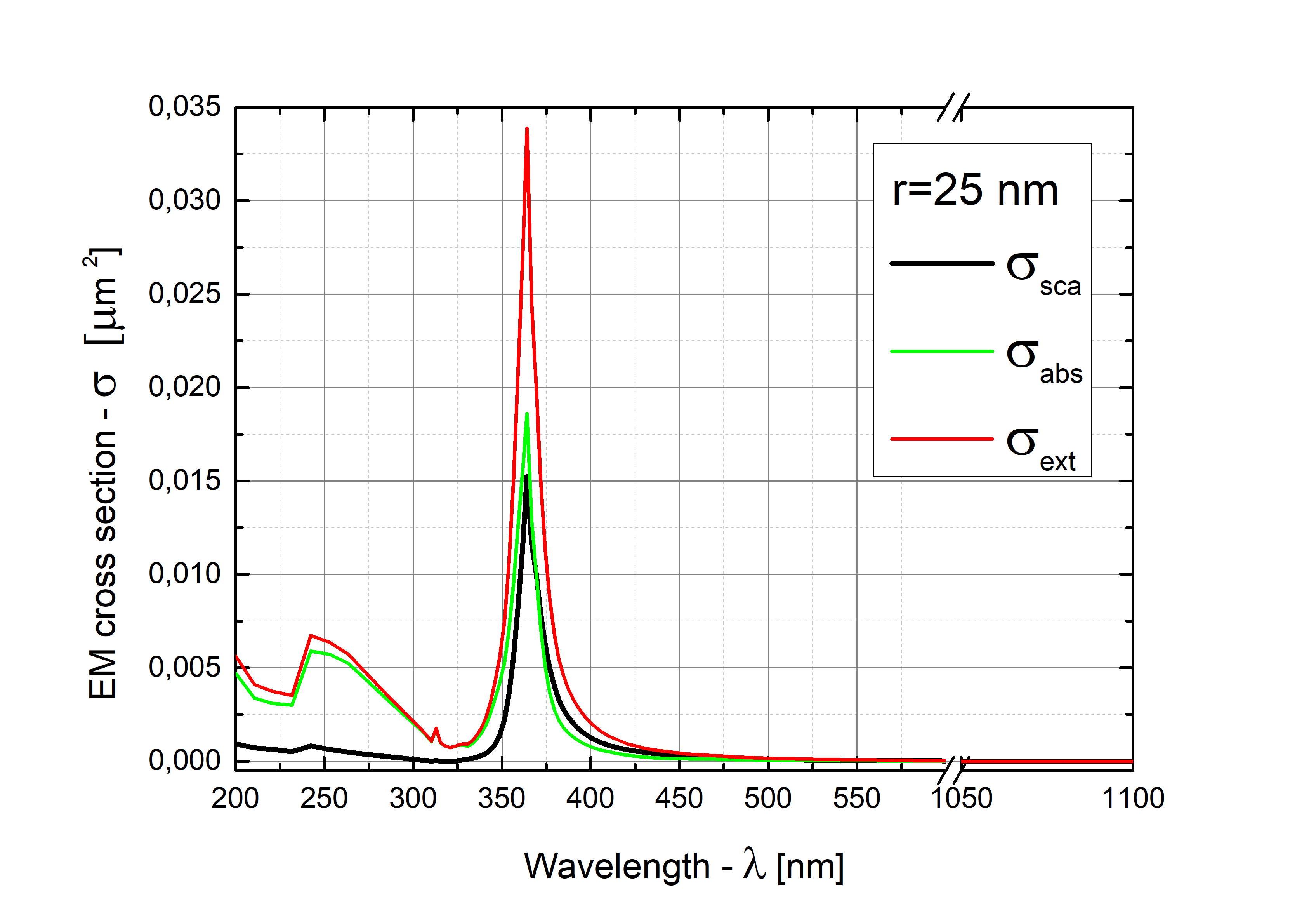}
		\caption{Effective cross sections}\label{Fig:2.3}
	\end{minipage}\hfill
	\begin{minipage}{0.5\textwidth}
		\centering
		\includegraphics[width=1.1\linewidth]{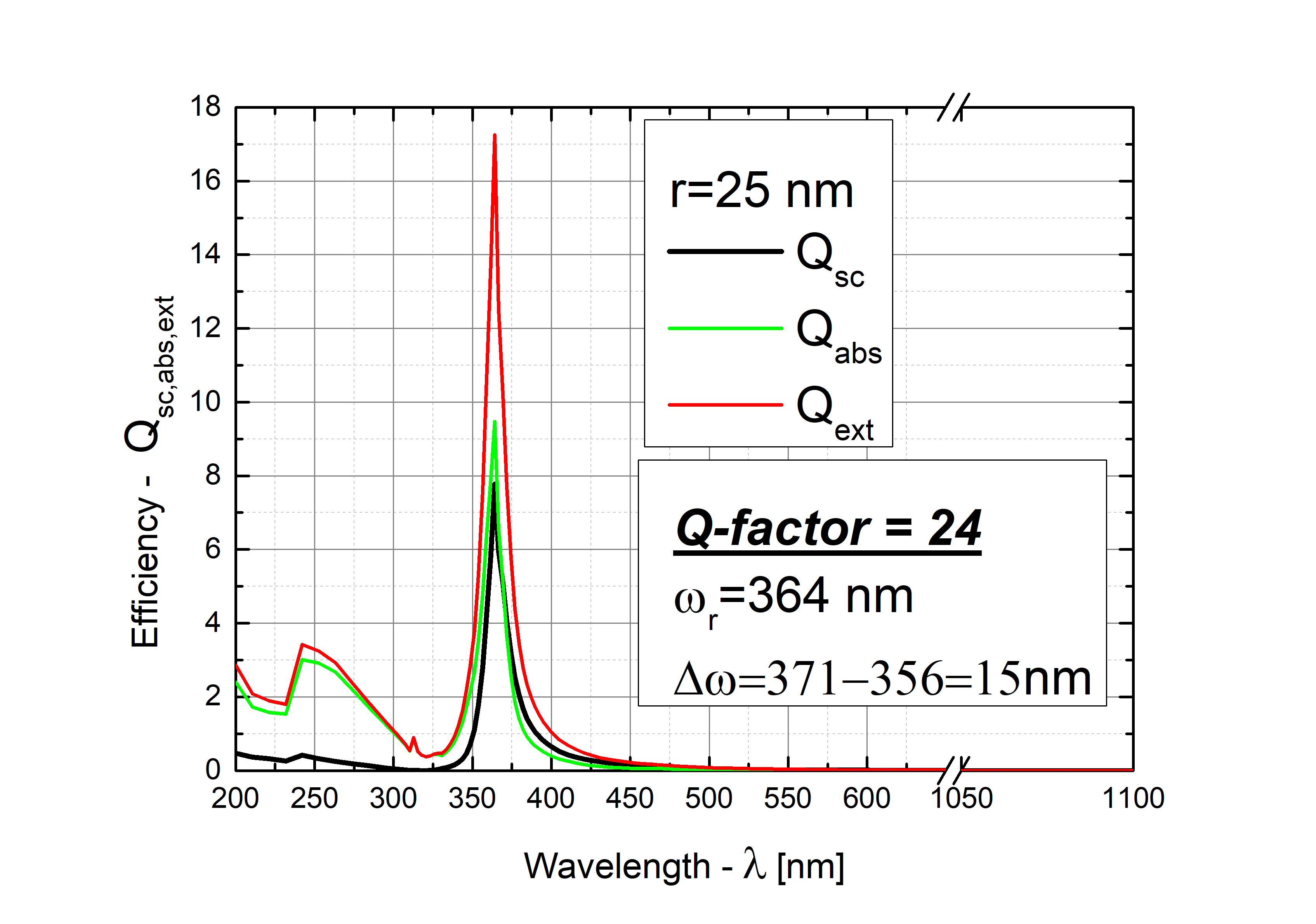}
		\caption{Efficiency of EM interception}
		\label{Fig:2.4}
	\end{minipage}
\end{figure}
\newpage 
The $\mathbb{Q}$-factor calculations are performed by extracting resonance frequency, $\omega_r$, and the full width half maximum, $\triangle$$\omega$ values from the extinction efficiency, $Q_{ext}$, graph. The quality factor is given as; 
\[\mathbb{Q}=\frac{\omega_r}{\triangle\omega}\]
\par 
Through parameter sweeping, sphere geometries for eight different size; scattering, absorption and extinction cross sections are calculated.
\begin{figure}[!htb]
	\begin{minipage}{0.5\textwidth}
		\centering
		\includegraphics[width=1.1\linewidth]{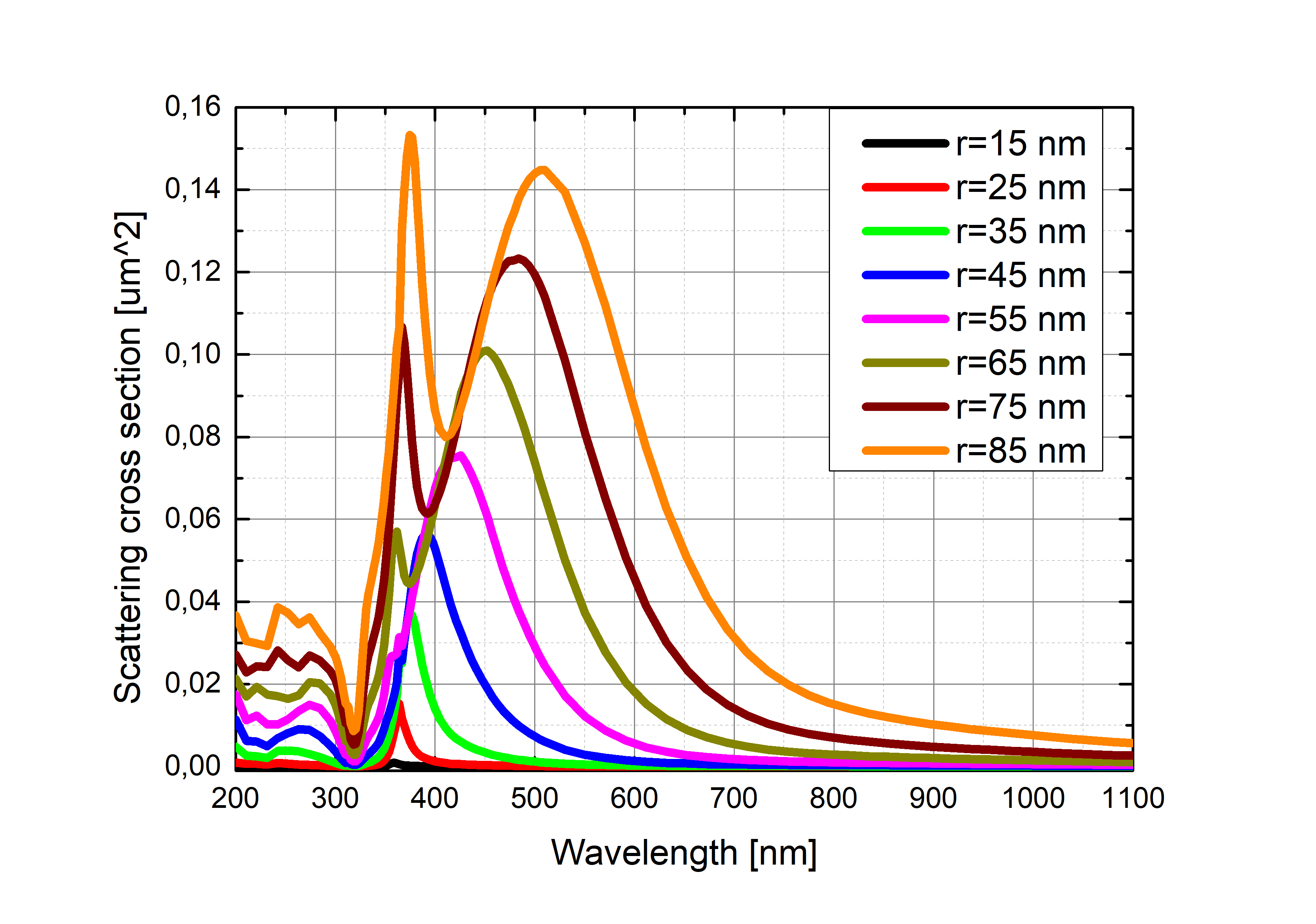}
		\caption{Scattering cross section\\ for different sizes}\label{Fig:sigmaSC_r}
	\end{minipage}\hfill
	\begin{minipage}{0.5\textwidth}
		\centering
		\includegraphics[width=1.1\linewidth]{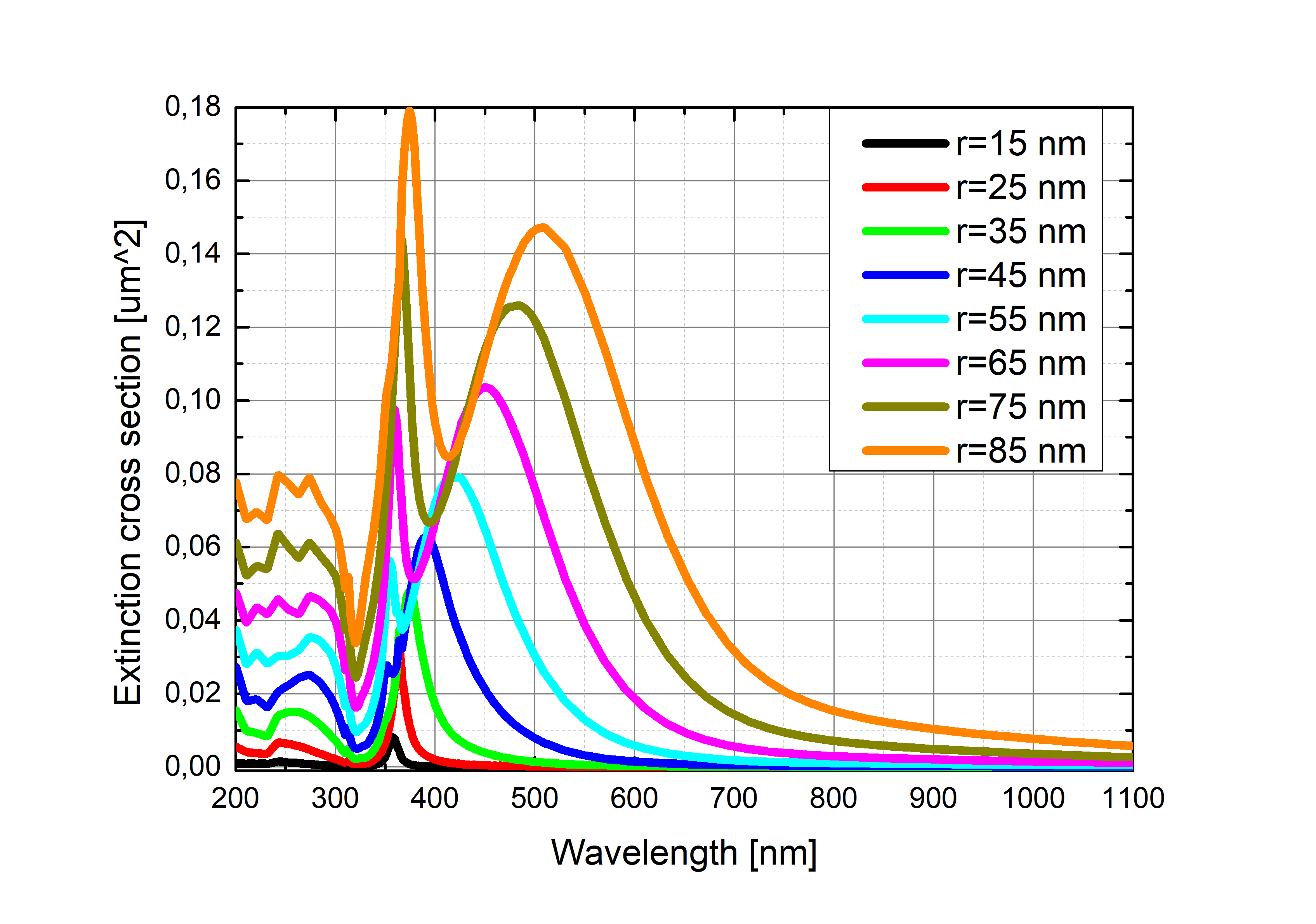}
		\caption{Extinction cross section\\ for different sizes}\label{Fig:sigmaEXT_r}
	\end{minipage}
\end{figure}
\begin{figure} [h!]
	\centering
	\includegraphics[scale=0.55]{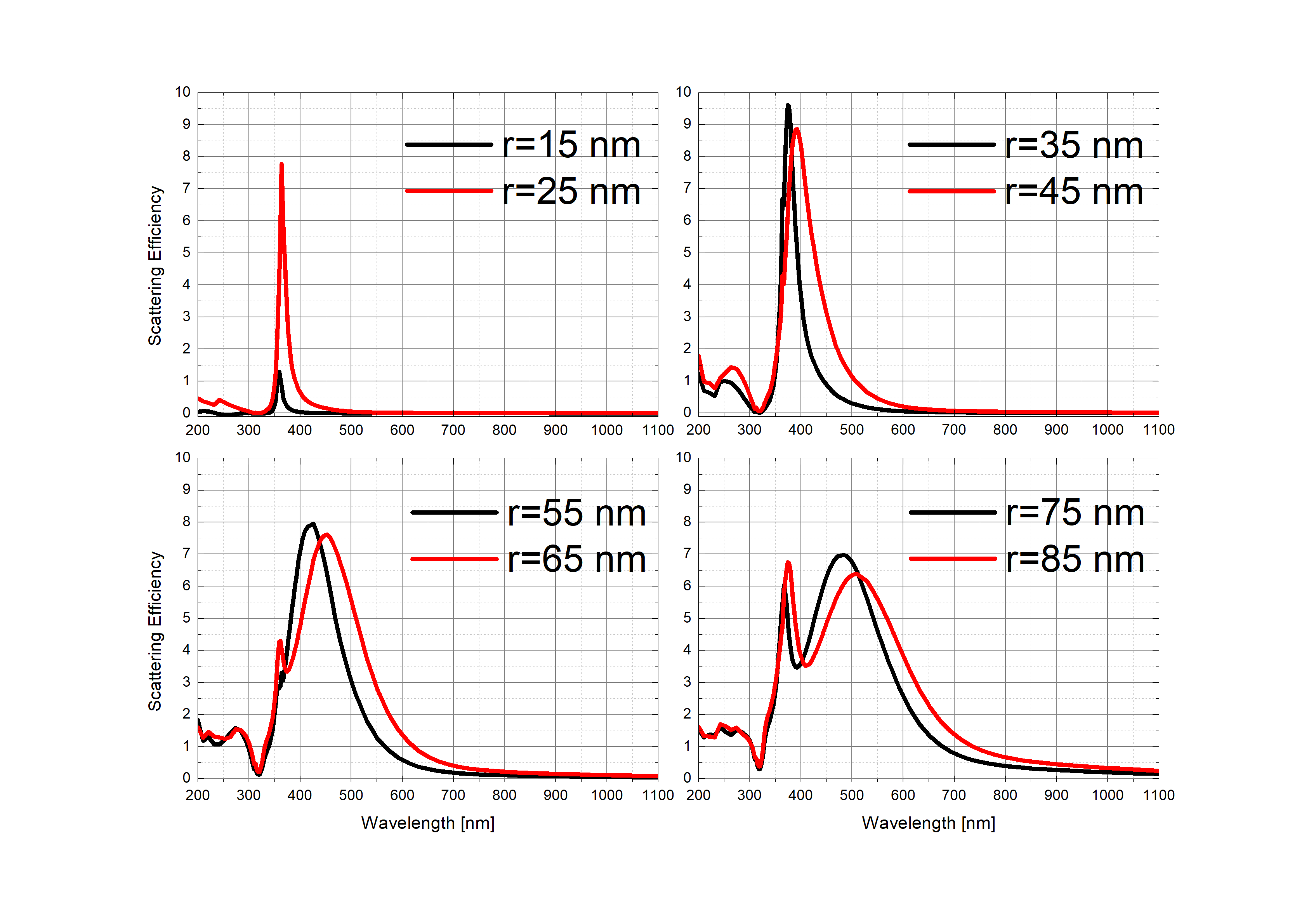} 
	\caption{Scattering Efficiencies}\label{Fig:Q_sc_r}
	\label{fig:Q_ext_r}
\end{figure}
\begin{figure} [h!]
	\centering
	\includegraphics[scale=0.55]{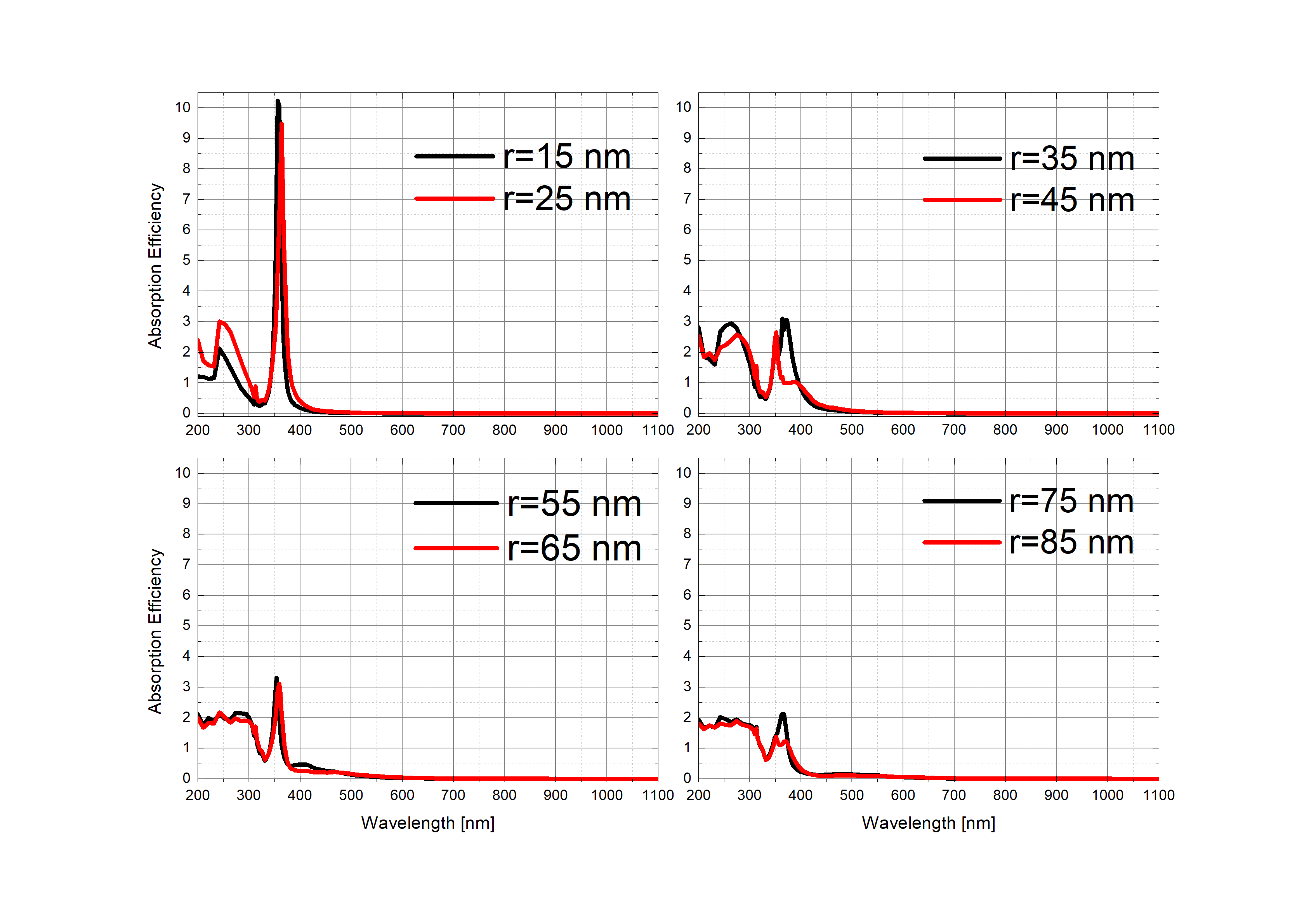} 
	\caption{Absorption Efficiencies}\label{Fig:Q_abs_r}
	\label{fig:Q_ext_r}
\end{figure}
\begin{figure} [h!]
	\centering
	\includegraphics[scale=0.55]{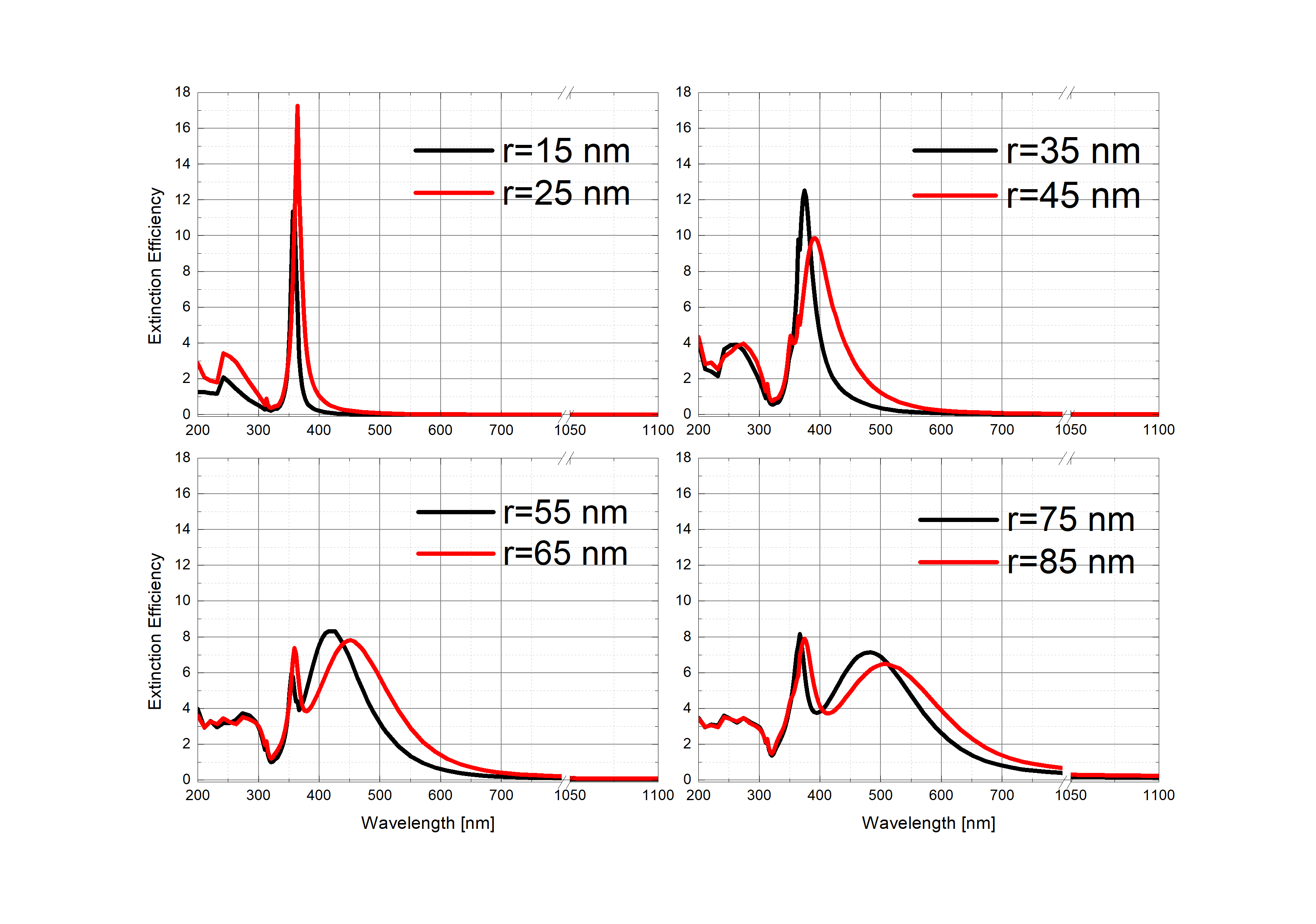} 
	\caption{Extinction efficiency comparison for different sizes}
	\label{fig:Q_ext_r}
\end{figure}

\begin{figure}[h!]
	\centering
	\includegraphics[width=0.8\linewidth]{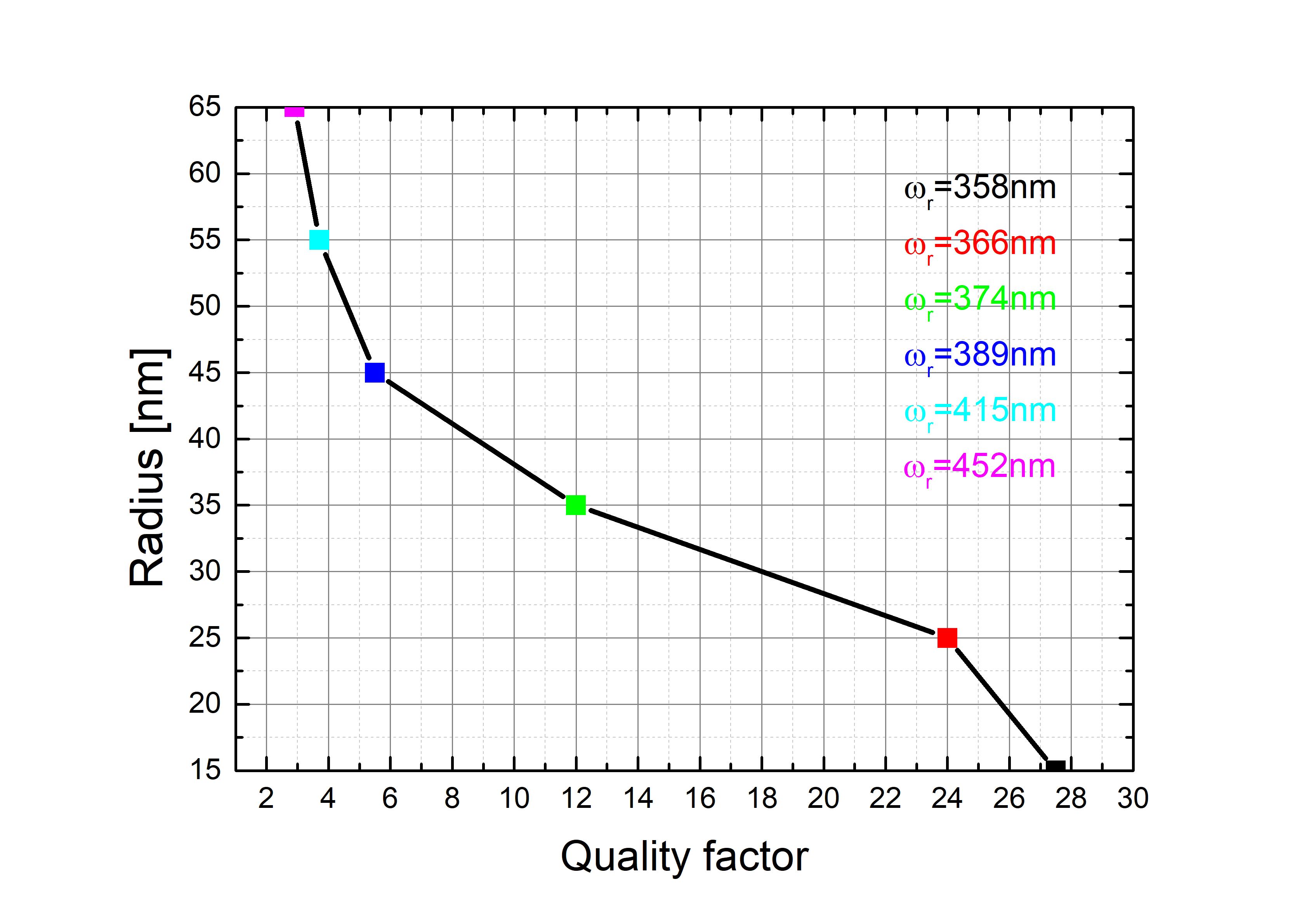}
	\caption{Quality factor values for different size particles at their resonance frequency}\label{Fig:2.5}
\end{figure}
Furthermore, the $\mathbf{E}$ field enhancement \cite{Krasnok2013}, $\delta^e$, and field localization parameter, $\hat{z}$, are measured by implementing point probes into the near field of the particle along the axis in polarization direction.
\newpage 
\begin{figure}[h!]
	\centering
	\includegraphics[width=0.8\linewidth]{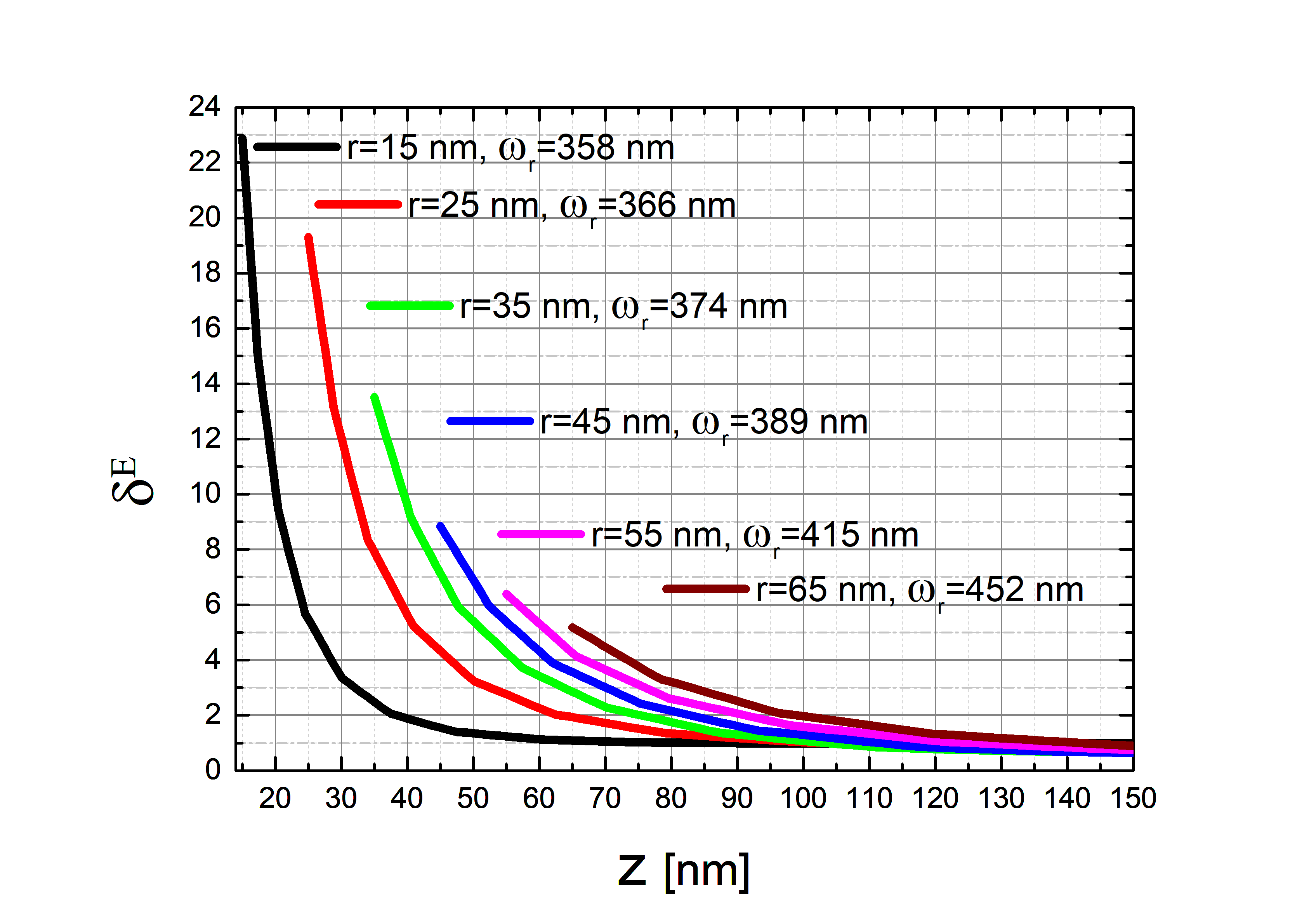}
	\caption{Field confinement in the vicinity of MNPs}\label{Fig:2.6}
\end{figure}
\newpage
The comparison of extinction efficiency, indicator of far field excitation coupling into smaller volumes, shows that there exists an optimized geometry for the best nanoantenna because its spectral response has to be narrow, as well.\par
The differences in the model with smaller geometries or the miscalculated data can be originated from the degree of approximation. In order to test this, as a primitive example; same case given above was calculated one more time but with a coarser mesh size which defines the discretization of finite elements. Applying only fine mesh onto the particle, the scattering cross section values, for high energies, started to change and in some values close to the zero they became to have negative values. These miscalculation effect is faced when getting smaller diameter in higher frequencies. The actual simulations are made mostly with extra fine mesh and it is seen that the unexpected data is meliorated. \par 
After that the effect of larger physical cross section on EM cross section was demonstrated in figure (\ref{Fig:sigmaSC_r}). As it was mentioned in \cite{Maier2011}, the spectral shift, towards to the infrared, has occured.\par

\cleardoublepage
\subsection{Scattering on nanoislands}
More realistic model is built to define a proper scattering model that gives explanatory results for the fabricated nanoparticles. By the help of information that we obtained from previous section above, an improved simulation model is developed for fabricated shapes. As we observed from the tilted SEM images, shape of the particles are hemispherical. Also, they are numerously formed. For the sake of reliability and clarity of the results, the model is built step by step in a progressive manner. Different contributions from substrate, shape, size and material will be included in this model. Therefore, each manipulation in the model is applied one at a time.\par 
At first, only change is to cut the sphere from its bottom. The particle suspended in midair which is surrounded by 200 nm \textit{perfectly matched layer} (PML). The mesh sizes are defined by considering the sharp edges and the particle's radius. For the nanoparticle domain, maximum element size is set as 2 nm and minimum element size is set as 1 nm. The remaning part in the physical domain is meshed as 'extremely fine' from the local mesh directory.  \par
Each model is computed with parametric sweep and proceeded around 26-27 hours with a powerful workstation. The computer has 20 cores and 40 processors with maximum 3.2 GHz speed and 192 GB random access memory (RAM). One model is computed with \%60 computer utilization. \par
It is obvious that the larger particle with 120 nm diameter has more effective cross section. The 30 nm diameter particle is absorption dominant while the larger one is dominant in the scattering regime. As it is expected, the quality factor of the resonance is decreased with the size increment. \par
The comparison of how a hemispheroid shape differs from a perfect sphere is given for 15 and 60 nm radius nanoantennas in figure \ref{fig:r15_sphereVShemi} and \ref*{fig:r60_hemi_air_comparison} . 
For the 30 nm size, the centre wavelength is shifted from 358 nm to 379 nm and the absorption efficiency becomes 17 while the perfect sphere has the value of 11. When the particle gets larger size, hemispheroid shape cause a shift in resonance wavelength and an increase in the total efficiency. For 120 nm, the centre wavelength is observed around 450 nm and the scattering efficiency is increased from 8 to 13. The extinction efficiencies shows that the smaller particle has the larger value which is quite meaningful because the extinction efficiency is an indicator of how strong the incoming light and the nanoantenna interact. The smaller one interacts with the incoming light more than the larger one but works in absorption dominant regime.
\begin{figure} [h!]
	\centering
	\includegraphics[scale=0.5]{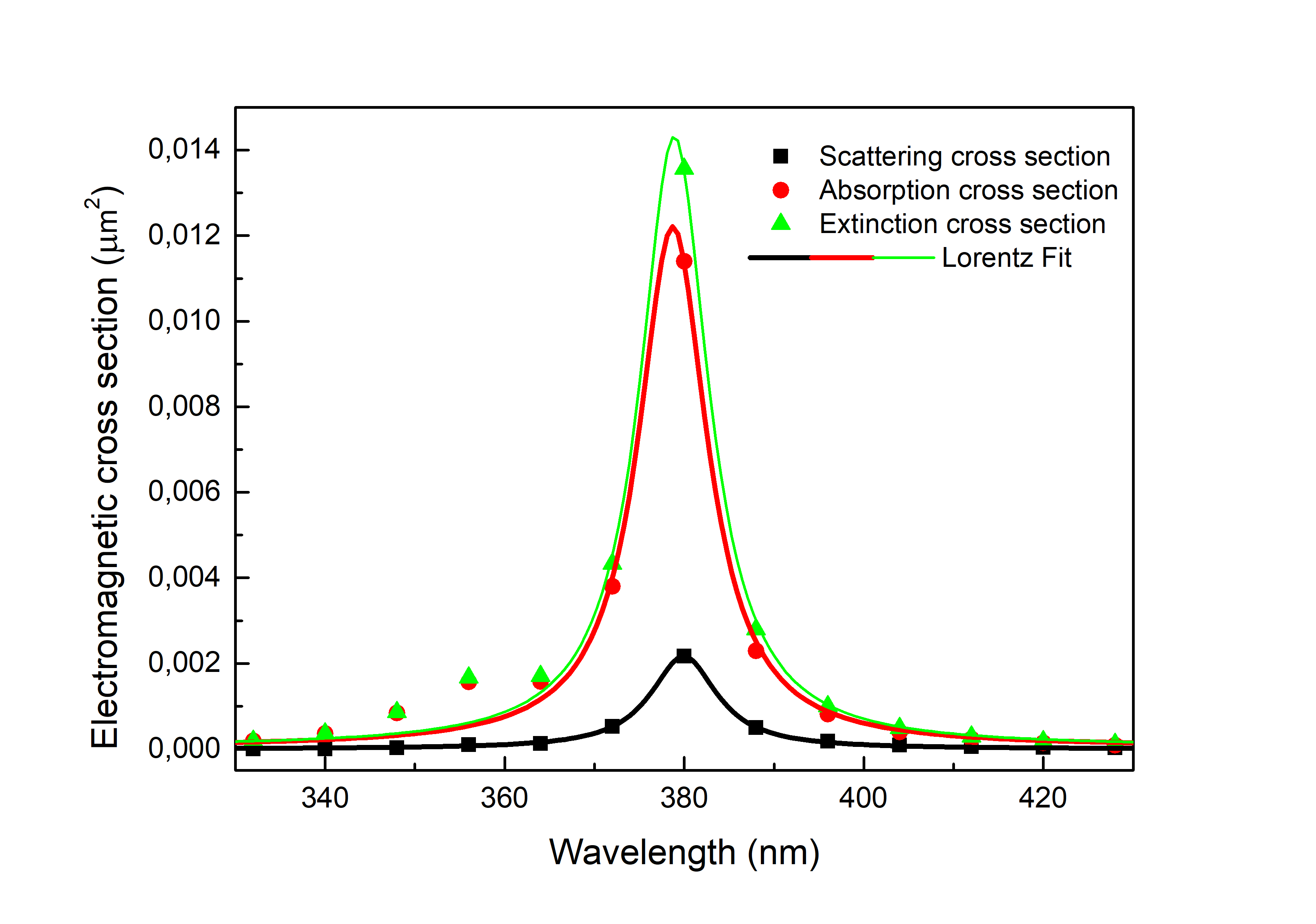} 
	\caption{Electromagnetic cross section values for a 15 nm radius hemisphere nanoantenna.}
	\label{fig:r15_hemi_air_CS}
\end{figure}
\begin{figure} [h!]
	\centering
	\includegraphics[scale=0.5]{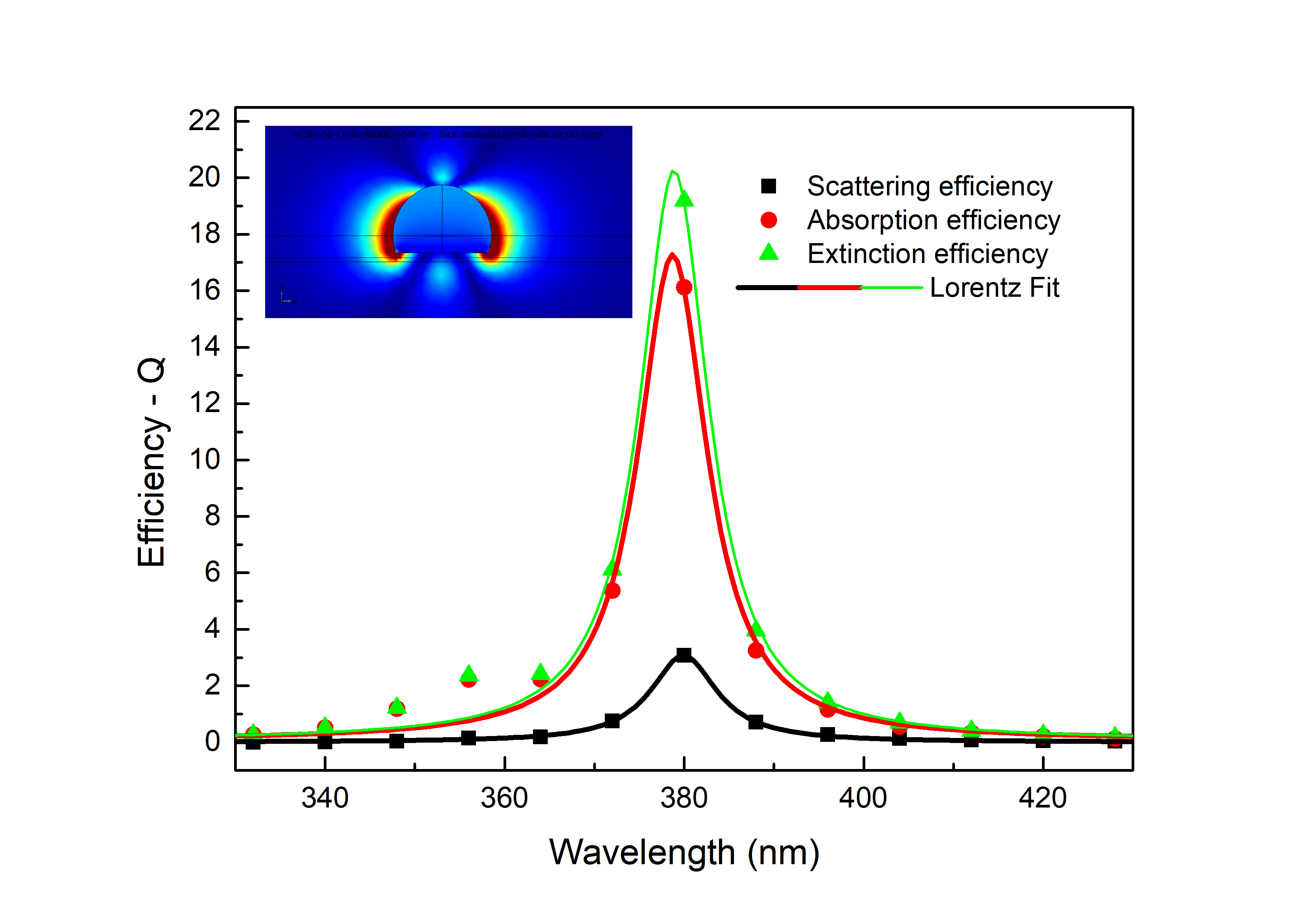} 
	\caption{Electromagnetic efficiencies for a 15 nm radius hemisphere nanoantenna.}
	\label{fig:r15_hemi_air_Q}
\end{figure}
\begin{figure} [h!]
	\centering
	\includegraphics[scale=0.5]{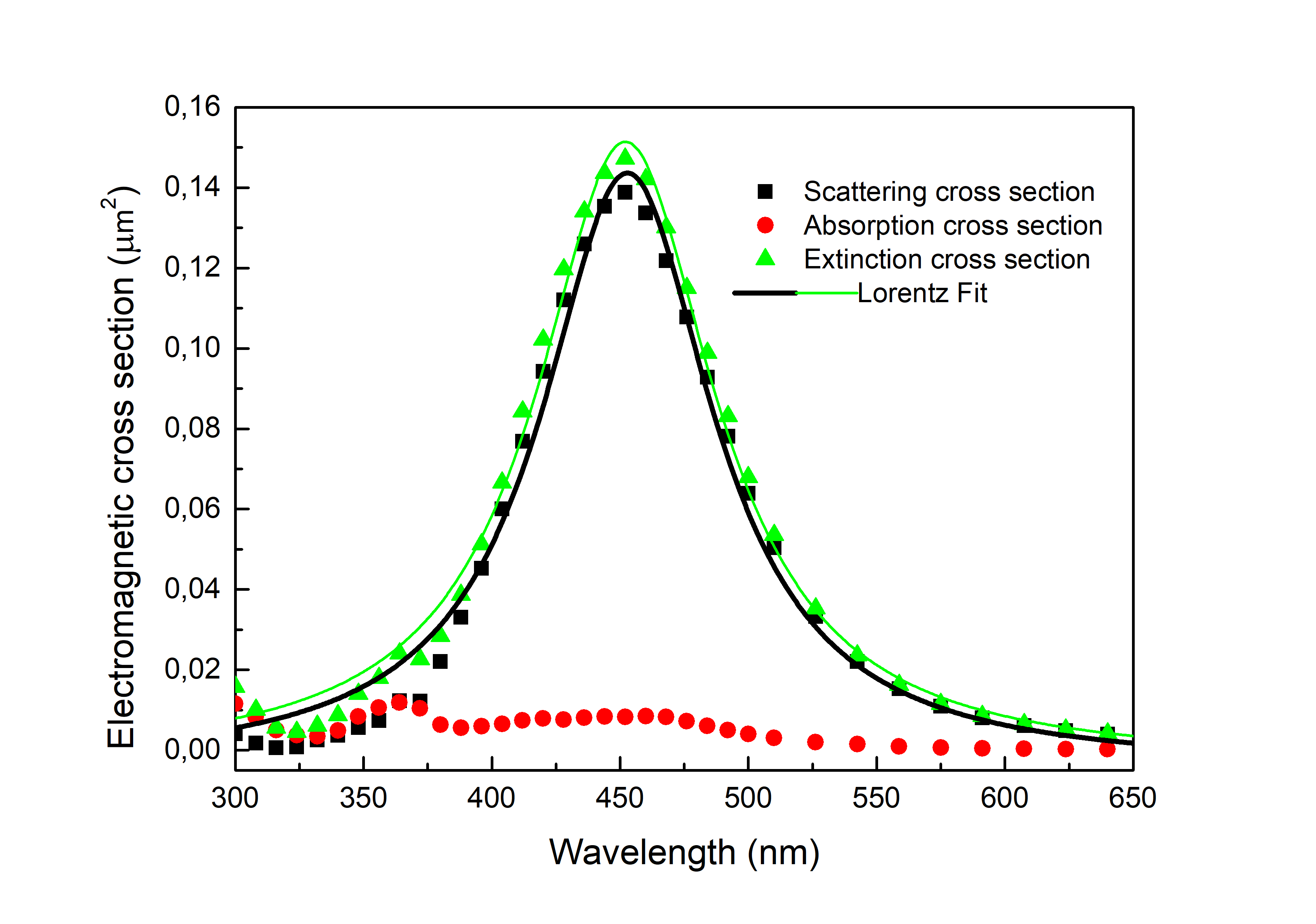} 
	\caption{Electromagnetic cross section values for a 60 nm radius hemisphere nanoantenna.}
	\label{fig:r60_hemi_air_CS}
\end{figure}
\begin{figure} [h!]
	\centering
	\includegraphics[scale=0.5]{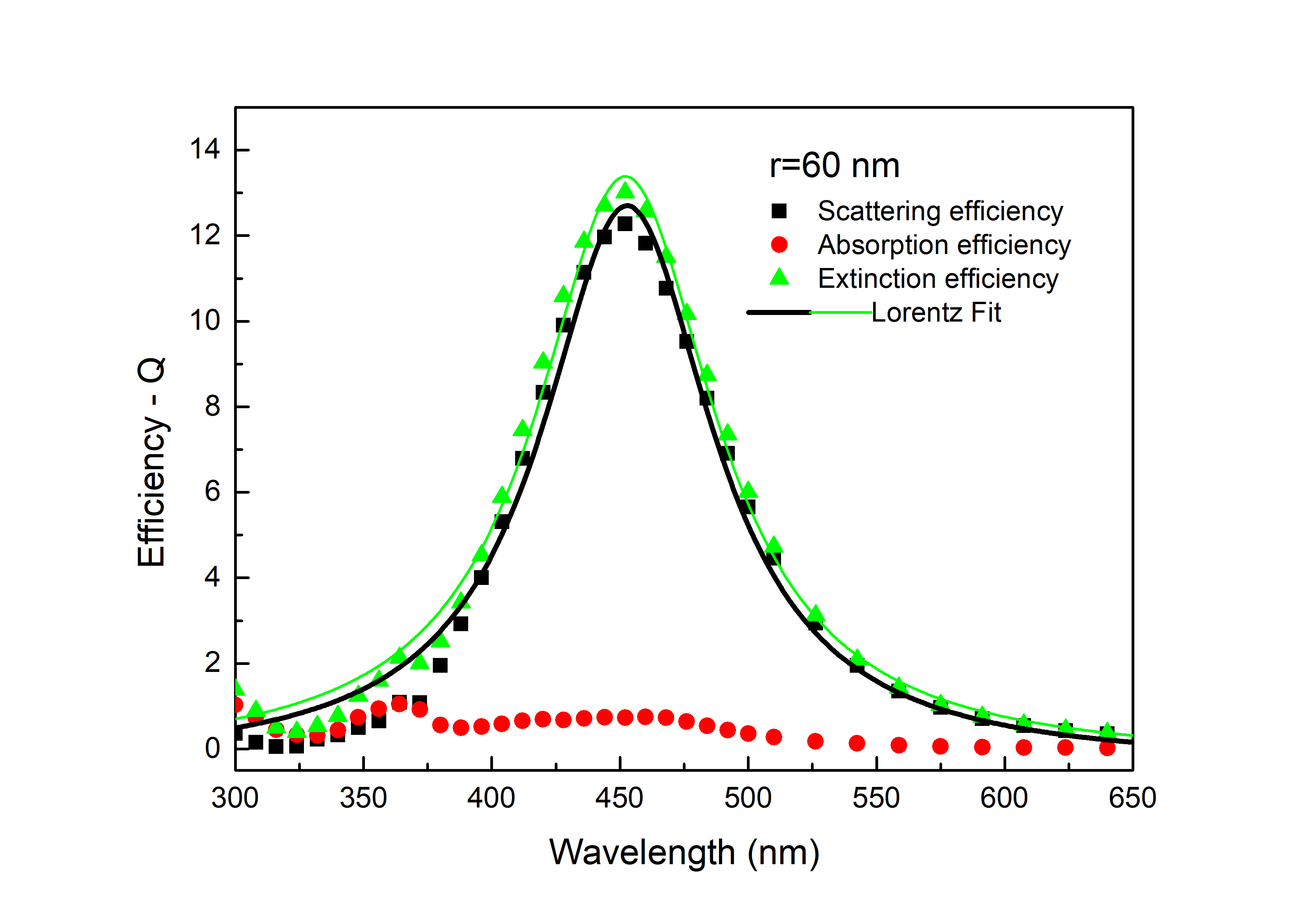} 
	\caption{Electromagnetic efficiencies for a 60 nm radius hemisphere nanoantenna.}
	\label{fig:r60_hemi_air_Q}
\end{figure}
\begin{figure} [h!]
	\centering
	\includegraphics[scale=0.5]{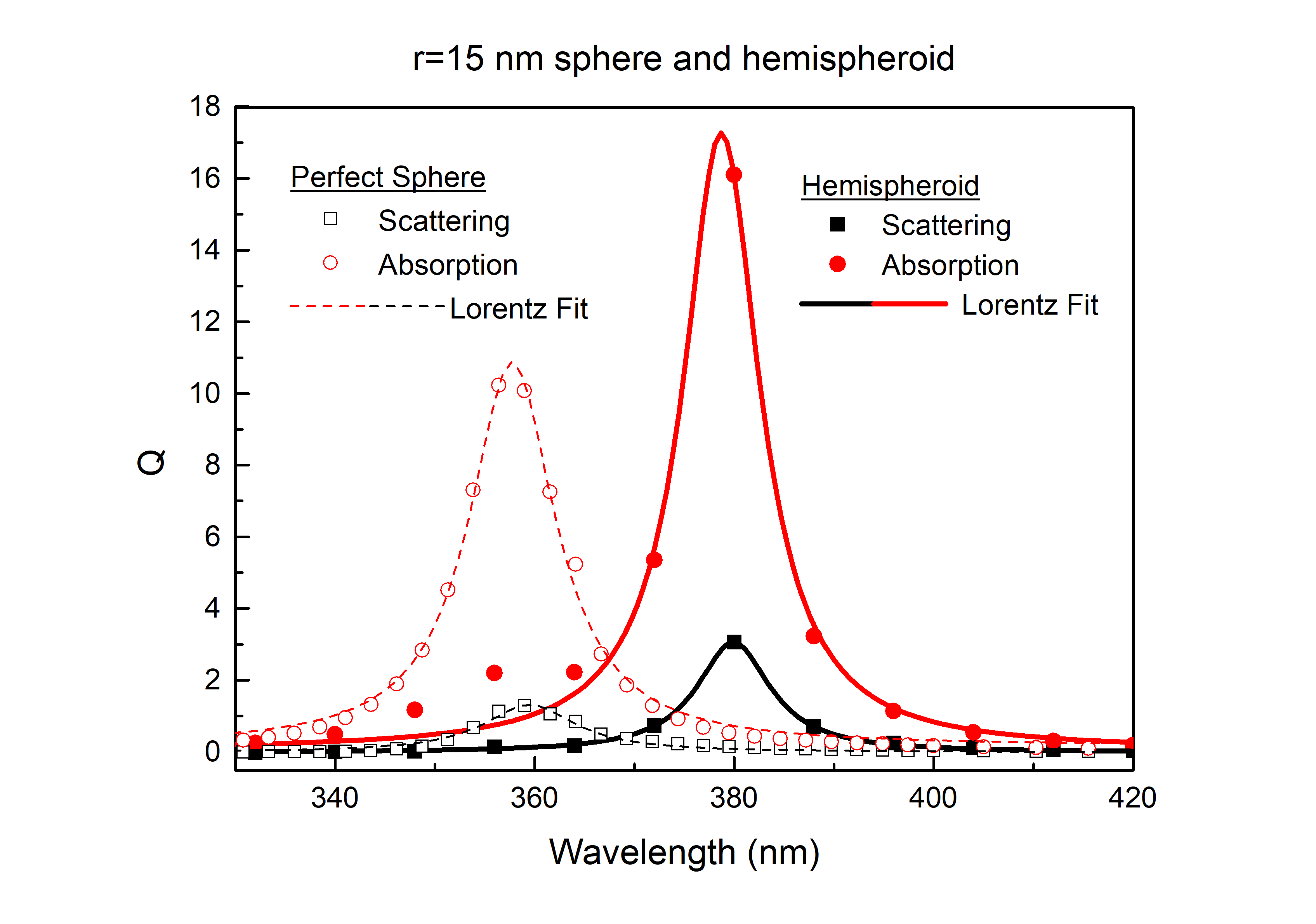} 
	\caption{Electromagnetic efficiency comparison of perfect sphere and hemisphere for 15 nm radius}
	\label{fig:r15_sphereVShemi}
\end{figure}
\begin{figure} [h!]
	\centering
	\includegraphics[scale=0.5]{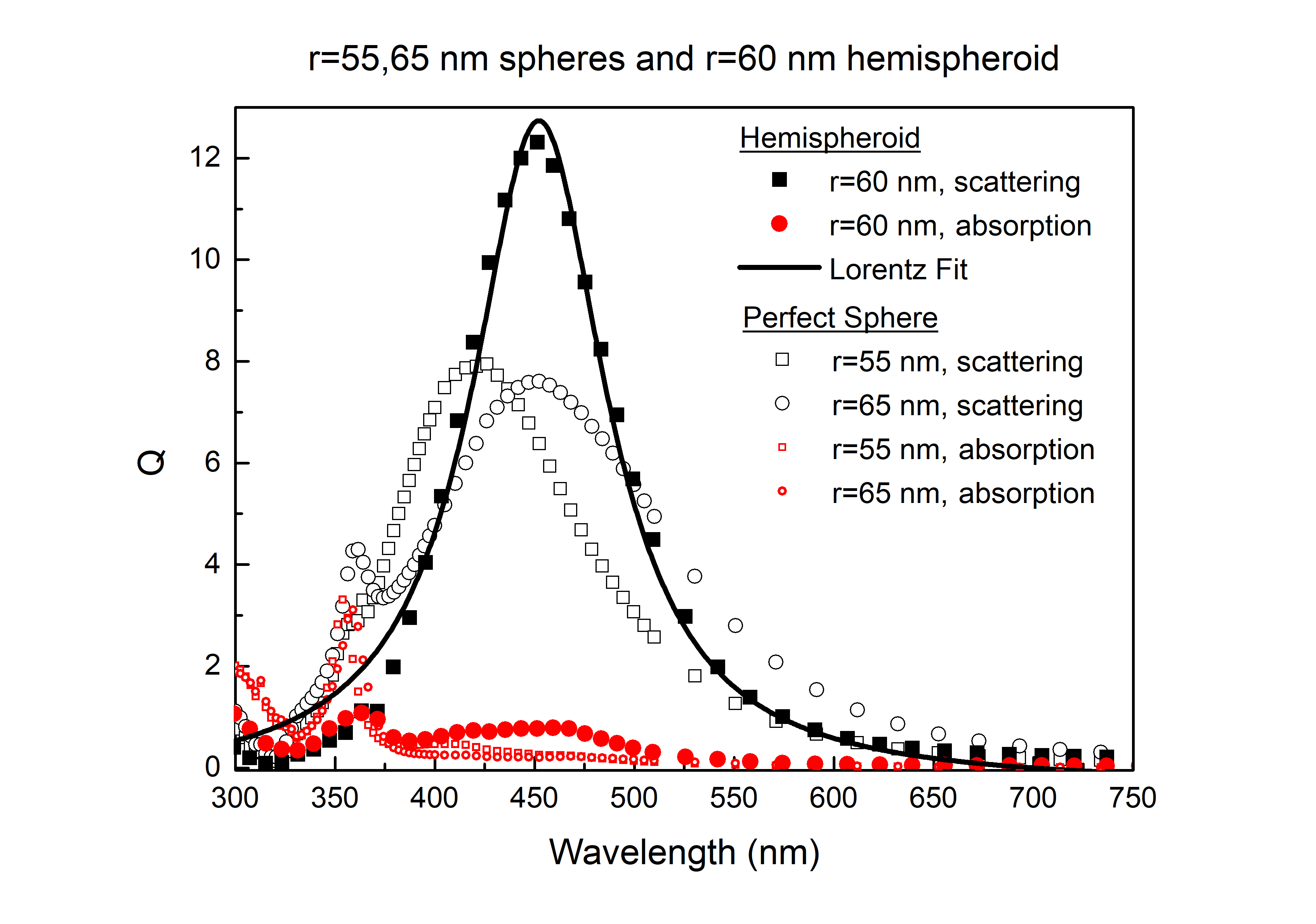} 
	\caption{Electromagnetic efficiency comparison of perfect sphere and hemisphere for 60 nm radius.}
	\label{fig:r60_hemi_air_comparison}
\end{figure} 
\cleardoublepage
\section{ Electromagnetic cross section and efficiency}
The subject is energy transfer from a far-field light source to a nano optical object. The energy flux of electromagnetic radiation is the energy flowing through a unit area in a unit time. It is given by Poynting vector, $\mathbf{S}$. \par  
The electric and magnetic fields are  $ \mathbf{E}=\mathbf{E(r)} e^{-i\omega t} $ and $\mathbf{H}=\mathbf{H(r)} e^{-i\omega t}$ where $\mathbf{E(r)}=\mathbf{E_r}$ and $\mathbf{H(r)=\mathbf{H_r}}$ are phasors.
\begin{align}
\mathbf{S}&=\mathbf{E} \times \mathbf{H}    \\
\mathbf{S}(t)&=\Re(\mathbf{E_r}e^{-i\omega t}) \times \Re(\mathbf{H_r}e^{-i\omega t}) \nonumber \quad \quad \quad \quad\quad \mbox{\textit{instantaneous Poynting vector}} \\
\mathbf{S}(t)&=\frac{1}{2}(\mathbf{E_r}e^{-i\omega t}+\mathbf{E^*_r}e^{i\omega t}) \times \frac{1}{2}(\mathbf{H_r}e^{-i\omega t}+\mathbf{H^*_r}e^{i\omega t}) \nonumber\\
\mathbf{S}(t)&=\frac{1}{4}(\mathbf{E_r}\times \mathbf{H^*_r} +\mathbf{E^*_r}\times \mathbf{H_r} + \mathbf{E_r}\times \mathbf{H_r} e^{-2i\omega t } + \mathbf{E^*_r} \times \mathbf{H^*_r}e^{2i\omega t} ) \nonumber\\
\mathbf{S}(t)&=\frac{1}{2}\Re(\mathbf{E_r}\times \mathbf{H^*_r}) + \frac{1}{2}\Re(\mathbf{E_r}\times \mathbf{H_r} e^{-2i\omega t } ) \nonumber
\end{align}
Time average of the instantaneous Poynting vector,
\begin{align*}
\langle \mathbf{S}(t) \rangle &= \frac{1}{T} \int_{0}^{T}\mathbf{S}(t) dt \\
\langle \mathbf{S}(t) \rangle &= \frac{1}{T} \int_{0}^{T} \left[ \frac{1}{2}\Re(\mathbf{E_r}\times \mathbf{H^*_r}) + \cancelto{0}{\frac{1}{2}\Re(\mathbf{E_r}\times \mathbf{H_r} e^{-2i\omega t }} ) \right] dt \\
\langle \mathbf{S}(t) \rangle &= \frac{1}{T} \int_{0}^{T} \frac{1}{2}\Re(\mathbf{E_r}\times \mathbf{H^*_r}) dt\\
\langle \mathbf{S}(t) \rangle &= \frac{1}{2}\Re(\mathbf{E_r}\times \mathbf{H^*_r})
\end{align*}
\[\rule{10cm}{1pt}\]
\begin{align}
\mathbf{S}_{inc}&= \frac{1}{2}\Re\{\mathbf{E}_{inc} \times {\mathbf{H^*}_{inc}} \} \\
\mathbf{S}_{sca}&= \frac{1}{2}\Re\{\mathbf{E}_{sca} \times {\mathbf{H^*}_{sca}} \} \\
\mathbf{S}_{ext}&= \frac{1}{2}\Re\{\mathbf{E}_{inc} \times {\mathbf{H^*}_{sca}} +\mathbf{E}_{sca} \times {\mathbf{H^*}_{inc}} \label{eq:2.16} \}
\end{align}
$\mathbf{S}_{inc}$ and $\mathbf{S}_{sca}$ are time-averaged electromagnetic power flows for incoming and scattered fields. Incoming and scattered fields also interfere with each other. This interaction causes an additional power expression which is being $\mathbf{S}_{ext}$ in equation (\ref{eq:2.16}).
\begin{align}
\mathbf{S}_{tot}&= \mathbf{S}_{inc}+\mathbf{S}_{sca}+\mathbf{S}_{ext}, \quad \quad \mbox{\textsc{Conservation of Energy}}\\
\mathbf{S}_{tot}&= \frac{1}{2}\Re\{\mathbf{E}_{tot} \times {\mathbf{H^*}_{tot}} \} \\
\mathbf{E}_{tot}&=\mathbf{E}_{inc}+\mathbf{E}_{sca}\quad \quad  \& \quad \quad  \mathbf{H}_{tot}=\mathbf{H}_{inc}+\mathbf{H}_{sca}
\end{align}
The crucial point is to obtain the energy values relevant to the particle. In the model (\ref{sec:2.1}), the surface, $s$, and volume, $v$, of the particle are defined as boundary and domain. And the integrations are performed as;
\begin{align}
P_{sca}&= \int_s \mathbf{S}_{sca} ds,  &\mbox{\textit{Scattered energy over the particle}}\\
P_{abs}&= -\int_s \mathbf{S}_{tot} ds,   &\mbox{\textit{Absorbed energy over the particle}}\\
P_{ext}&= -\int_s \mathbf{S}_{ext} ds, &\mbox{\textit{Total energy removed from the incident field}}
\end{align}

The scattering and the absoption processes remove energy from the incident field. Throughout the interaction of light with the nanoparticle, incoming fields experience an obstacle larger than the particle's cross sectional area. The electromagnetic cross sections are given as \cite{Novotny2006};
\begin{align}
\sigma_{sc}= \frac{P_{sca}}{|\mathbf{S}_{inc}|}, \quad \quad 
\sigma_{abs}= \frac{P_{abs}}{|\mathbf{S}_{inc}| }, \quad \quad
\sigma_{ext}= \frac{P_{ext}}{|\mathbf{S}_{inc}|}=\sigma_{sc}+\sigma_{abs}
\end{align}
As a criterion of how the interception area has a larger value than the actual cross sectional area, the efficiencies are given as;
\begin{align}
Q_{sc}=\frac{\sigma_{sc}}{area}, \quad \quad 
Q_{abs}=\frac{\sigma_{abs}}{area}, \quad  \quad 
Q_{ext}=\frac{\sigma_{ext}}{area}=Q_{sc}+Q_{abs}
\end{align}

\[ Q_{ext} \quad \quad \implies \mbox{\textit{Strong interaction with the incoming fields}}    \]
\cleardoublepage
\subsection{Computational investigation of the quantum emitter}
\paragraph*{Fermi's Golden Rule}
The emission rate for a single isolated atom is expressed as\cite{Novotny2006};
\begin{align}
\gamma= \frac{2\pi}{\hbar^2} |\langle f| \hat{H}_I |i \rangle|^2 \delta(\omega_i-\omega_f)
\end{align}
$\delta(\omega_i-\omega_f)$ expresses the transition from initial to final state. $\hat{H}_I$ is the interaction Hamiltonian between atom and photon. 
Since we have a numerous of final states, the above expression should be sum over all final states;
\begin{align}
\gamma= \frac{2\pi}{\hbar^2} \sum_{f} &|\langle f| \hat{H}_I |i \rangle|^2 \delta(\omega_i-\omega_f) \\
&|\langle f| \hat{H}_I |i \rangle|^2= \langle i|\hat{\mu}.\hat{E}|f\rangle \langle f|\hat{\mu}.\hat{E}|i\rangle
\end{align}
\begin{align*}
|f\rangle \langle f| \quad  \rightarrow \quad \textit{projection operator which is a scanner for f states}
\end{align*}
In case of infinitely many photons, the final states become continuous. The number of final single-photon states are called \textbf{photonic density of states} (LDOS), $\rho(r_0,\omega_0)$, where $r_0$ is the position of the two-level quantum emitter. \par 
Partial LDOS is defined as;
\begin{align}
\rho_\mu(r_0,\omega_0)=\frac{6\omega_0}{\pi c^2}\left[ \mathbf{n}_\mu. Im\{ \overleftrightarrow{G}(r_0,r_0,\omega_0) \} .\mathbf{n}_\mu\right]
\end{align} \\
and the total density of photonic state is the total electromagnetic modes per unit volume and unit frequency at a single location, $r_0$;

For free space;
\begin{align}
\rho_0= \frac{\omega_0}{\pi^2 c^3} \quad  \textit{and} \quad \gamma_0=\frac{\omega_0^3 |\vec{\mu}|^2}{3\pi\epsilon_0 \hbar c^3}
\end{align}
where \textbf{$\mu$}$=\langle f |\hat{\mu} | i \rangle $ is the transition dipole matrix element.\par 

Emission rate is expressed as;
\begin{align}
\boxed{\gamma= \frac{2 \omega }{3 \hbar \epsilon_0 } |\mu|^2 \rho_\mu(\vec{r}_0, \omega_0)}
\end{align}

Partial local density of states, $\rho_\mu$, represent final photon states. According to Fermi's Golden Rule and fluctuation-dissipation theorem, continuous modes should be taken into account in order to investigate the quantum mechanical description of an inhomogeneous lossy medium. It is also possible to consider the lossy medium as a continuous photonic reservoir.\par 

The quantum emitter is a radiating electric dipole. Normalized quantum mechanical decay rate is equal to the normalized power radiated from a point dipole in an inhomogeneous environment \cite{Novotny2006}. The radiated power has to be normalized with the power in the absence of the inhomogeneous environment.  \par
\begin{align}
\frac{\gamma}{\gamma_0}=\frac{P}{P_0} 
\end{align}
By starting with the classical radiation power rate of an electric dipole, it is possible to survey the effects of the inhomogeneous environment's contribution to dipole emission. These results reflect the inhomogeneous environment's response and do not relevant to the properties of hBN defects at all. The normalized values emhasize that the dipole moment term is cancelled out while normalizing it with corresponding free-space values. Then, local density of photonic state, quantum efficiency, radiative and non-radiative decay rates can be calculated.\par

The power radiated from the point dipole is integrated over the boundaries of a small mesh box (nonphysical) surrounding the point dipole for loss decay rate calculation and the outer surface of entire simulation domain for radiation decay rate. Another simulation is run by only removing the MNP from the same model and power values are normalized.\par 

Radiative and non-radiative decay rate contributions are computed just by changing size and position of the emitter. The non-radiative part expresses photon losses. Computations are made by wavelength sweeping since the spectral response of inhomogeneous environment is investigated. In order to observe two different regimes, 30 nm and 120 nm diameter particles are modelled and compared for each case. The \textbf{Purcell factor} is known for the weak coupling regime as, \par 
\begin{equation} \label{eq:Purcell}
\ F= \frac{3}{4\pi^2} (\frac{\lambda_0}{n})^3 \frac{Q}{V}  \  
\end{equation}

The term $\lambda_0/n $ is the resonance wavelength for the material with refractive index of $n$. Quality factor (Q) and the mode volume (V) are tools for manipulation of any quantum emitter located in a resonant medium. Mode volume, V, is an electromagnetic quantity that measures the local density of photonic states (it is known as a physical volume for dielectric cavities). The real and imaginary parts of V are responsible for on and off resonances. Re(V) emphasizes the on resonances therefore supports the radiative part while the Im(V) emphasizes the off resonances. In other words, the imaginary part is corresponded by the non-radiative LDOS and the real part is corresponded with radiative LDOS.\par

The quantum emitter can be located at any point around a particle. It is important to find LDOS of that point where quantum emitter is located. \par

\begin{figure} [h!]
	\centering
	\includegraphics[scale=0.48]{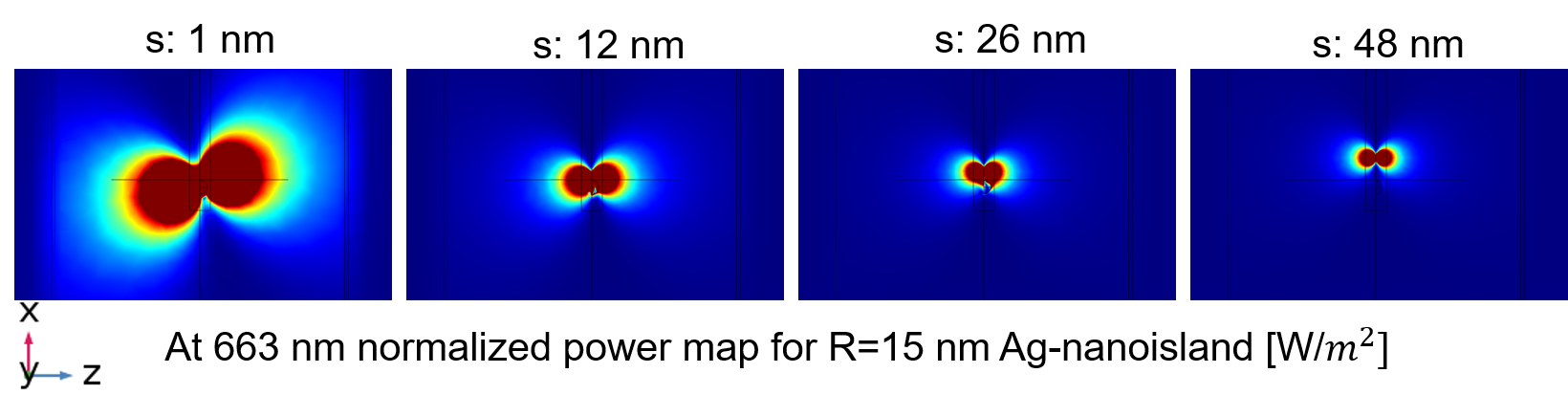}
	\includegraphics[scale=0.4]{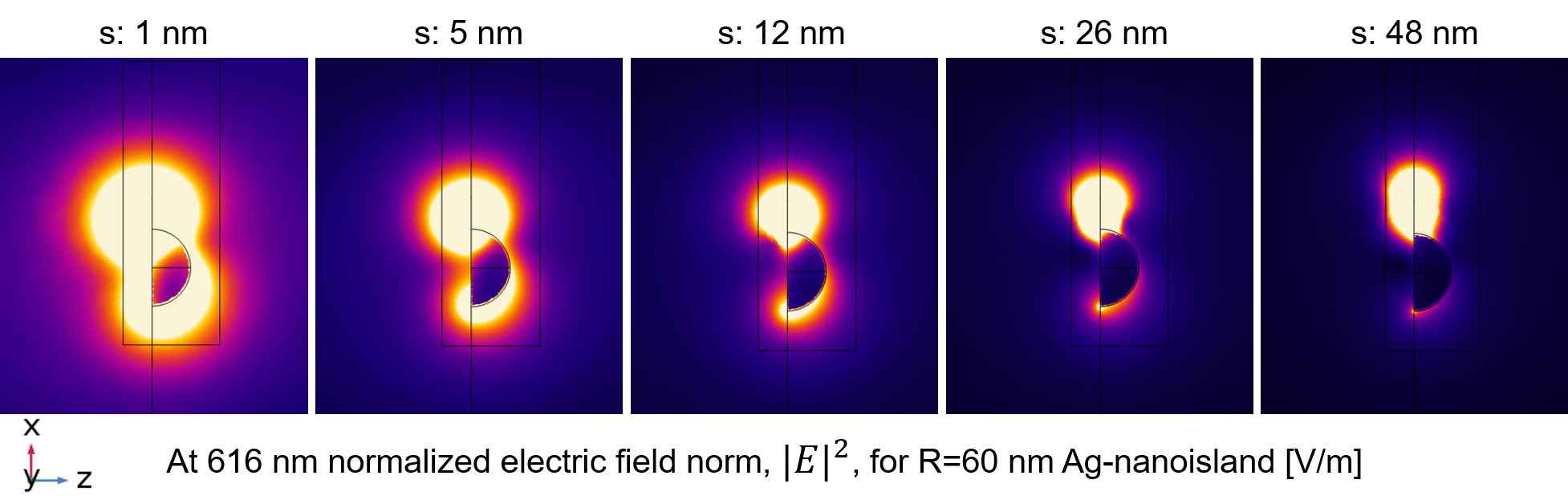}
	\caption{The dipole emitter getting closer to the nanoantenna causes shift in dipole orientation of the QE-AgNP system. This confirms the experimental results. Color maps: Top: Power values. Bottom: Electric field norm for the quantum emitter-nanoantenna hybrid system.}
	\label{colormaps}
\end{figure}

\end{suppinfo}

\cleardoublepage
\bibliography{mybib}
\end{document}